\documentclass[twocolumn]{aastex61}
\usepackage{color}
\usepackage{graphics}
\usepackage{amssymb,amsmath}
\usepackage{amsmath}
\usepackage{psfrag}
\usepackage{graphicx}

\usepackage{float}
\usepackage[caption = false]{subfig}

\usepackage{color}
\usepackage{graphics}
\usepackage{amssymb,amsmath}
\usepackage{amsmath}
\usepackage{psfrag}
\usepackage{graphicx}

\newcommand{\f}{\frac}

\newcommand{\be}{\begin{equation}}      
\newcommand{\ee}{\end{equation}}      
      
\newcommand{\bef}{\begin{figure}}      
\newcommand{\eef}{\end{figure}}      
\newcommand{\bea}{\begin{eqnarray}}    
\newcommand{\eea}{\end{eqnarray}}

\def\bse{\begin{subequations}}
\def\ese{\end{subequations}}

\def\lsim{\raise 0.4ex\hbox{$<$}\kern -0.8em\lower 0.62ex\hbox{$\sim$}} 
\def\gsim{\raise 0.4ex\hbox{$>$}\kern -0.7em\lower 0.62ex\hbox{$\sim$}}

\def\f0N{f_0^{(N)}}
\def\bec{\begin{center}}
\def\eec{\end{center}}

\received{\today}
\revised{yyyy}
\accepted{zzzz}
\submitjournal{ApJ}
\shorttitle{Transient spiral arms from far out of equilibrium gravitational
  evolution}
\shortauthors{Benhaiem et al.}

\begin{document}

\title {Transient Spiral Arms from Far Out-of-equilibrium Gravitational Evolution} 
 
  \author{David Benhaiem} 
  \affil{Istituto dei Sistemi
  Complessi, Consiglio Nazionale delle Ricerche, Via dei Taurini 19,
  I-00185 Roma, Italia} 
  \author{Michael Joyce} \affil{Laboratoire de Physique
  Nucl\'eaire et de Hautes \'Energies, UPMC IN2P3 CNRS UMR 7585,
  Sorbonne Universit\'es, 4, place Jussieu, 75252 Paris Cedex 05,
  France} 
  \author{Francesco Sylos Labini} \affil{Centro Studi e
  Ricerche Enrico Fermi, Via Panisperna 00184, Roma, Italia }
\affil{Istituto dei Sistemi Complessi, Consiglio Nazionale delle
  Ricerche, Via dei Taurini 19, I-00185 Roma, Italia}
\affil{INFN Unit Rome 1, Dipartimento di Fisica, Universit\'a di
  Roma Sapienza, Piazzale Aldo Moro 2, 00185 Roma, Italia }

%\maketitle %\maketitle must follow title, authors, abstract and \pacs

\date{\today}

\begin{abstract}
{We describe how a simple class out-of-equilibrium, rotating,   
and asymmetrical mass distributions evolve under 
their self-gravity to produce a quasi-planar spiral structure  surrounding a virialized core, qualitatively resembling a spiral galaxy.  The spiral structure is transient, but can survive tens of dynamical times, and further reproduces qualitatively  noted features of spiral galaxies such as the predominance of trailing two-armed spirals and large pitch angles.  As our models are highly idealized, a detailed comparison with observations is not appropriate, but generic features of the velocity  distributions can be identified to be the potential observational signatures of such a mechanism. Indeed, the mechanism leads generically to a characteristic transition from predominantly rotational  motion, in a region outside the core, to radial ballistic  motion in the outermost parts. Such radial motions are excluded in our Galaxy up to 15 kpc,  but could be detected at larger scales in the future by GAIA. We explore the apparent motions seen by external  observers of the velocity distributions of our toy galaxies, and find that it is difficult to distinguish them from those of a rotating disk with sub-dominant radial motions at levels typically inferred from observations. These simple models illustrate the possibility that the observed apparent motions of spiral galaxies might be explained by non-trivial non-stationary mass and velocity distributions  without invoking a dark matter halo or modification of Newtonian gravity.  In this scenario the observed phenomenological  relation between the centripetal and gravitational acceleration of the visible baryonic mass could have a simple explanation.
}
\end{abstract}


%%%%%%%%%%%%%%%%%%%%%%%%%%%%%%%%%%%%%%%%%%%%%%%%%%%%%%%%%%%%%%%%%%%%%%%

\section{Introduction} 

The arms of spiral galaxies are one of the most striking and
remarkable features of the visible universe revealed by
astronomy. They have been the subject of much study, both
observational and theoretical, over many decades.  Several competing
theories have been advanced to explain their physical origin, but no
single one has emerged definitively as the correct framework {(see,
  e.g., \cite{Dobbs_Baba_2014})}. Understanding of their motions is
of particular importance because it is the observed apparent {(i.e.,
  on the line of sight --- LOS)} motions in the outer parts of spiral
galaxies that have led to the supposition that much of the
gravitating matter in them is not visible \citep{Rubin_1983}.  These
same motions have led also to alternative scenarios involving strong
modifications of Newtonian gravity \citep{Milgrom_1983}.  In this
paper we show how mass distributions  {\it qualitatively} resembling
those of the visible components of spiral galaxies can result from the
far out-of-equilibrium dynamics of purely self-gravitating systems,
starting from a class of {very simple idealized} initial conditions.
{We study in particular the generic features of the velocity 
distributions of the structures produced by this mechanism,
and consider their qualitative compatibility with observations
of motions in spiral galaxies.}

Our approach is different from standard theoretical ones, in which
spiral structure arises by {perturbation} (internal or external) of an
equilibrium system, and the large-scale motions are modeled assuming
a stationary mass distribution. Indeed, our study illustrates how, for
intrinsically non-stationary models, the relation between {apparent
motions and the associated mass distribution can be completely
different from that in stationary models}.  In particular, we show that the
observation of a non-Keplerian {rotation curve} in the outer part of
such a structure does not necessarily require the existence of an
extended dark matter halo {or modification of Newtonian gravity}, 
and could instead be consistent with
non-axisymmetric radial motion of weakly bound and unbound mass.

{We note that, because our models involve only Newtonian
  gravity, the physics we describe could potentially be applicable to
  astrophysical systems of very different natures and sizes --- to
  dwarf galaxies that are inferred from their motions to be even
  more dark-matter dominated than spirals (see, e.g., \cite{Combes_2002});
 to protoplanetary disks, which have been revealed in observations in
the last couple of years to have spiral-like structure (see, e.g., \cite{Christiaens_etal_2014}); 
or even possibly to circumplanetary disks, whose existence is still
inconclusive  (see, e.g., \cite{Ward_Canup_2010}).  In a forthcoming work
  \citep{benhaiem+joyce+syloslabini_2017} that is 
  complementary to this paper, we
 will describe the physical
mechanism 
in much greater detail, using both for a broad range of initial conditions and
also numerical simulations with larger particle numbers.}

The class of models we consider as initial conditions consists of
asymmetrical and isolated self-gravitating clouds with some angular
momentum.  The dynamics of isolated self-gravitating systems from 
out-of-equilibrium initial conditions has been extensively studied for
several decades
\citep{Henon_1973,vanalbada_1982,aarseth_etal_1988,david+theuns_1989,aguilar+merritt_1990,theuns+david_1990,boily_etal_2002,barnes_etal_2009}. Broadly
speaking, such systems relax quite efficiently to virial equilibrium,
i.e., on time scales of the order of a few times the characteristic dynamical
time. Early studies showed that spherical configurations with little
isotropic velocity dispersion (i.e. sub-virial, with an initial virial
ratio {$b>-1$}) could produce equilibrated structures resembling
elliptical galaxies, with  a surface brightness notably close to the
observed de Vaucouleurs law \citep{vanalbada_1982}.  One generic
feature of such sub-virial collapses (for {$b \ge -0.5$}) is that they
lead to the ejection of some of the initial mass (see,
e.g., \cite{ejection_mjbmfsl,syloslabini_2012,syloslabini_2013}): the
strong contraction of the initial configuration leads to a rapidly
varying mean-field, which causes particle energies to 
also rapidly
vary, leaving some of them weakly bound and others with positive
energy.  Two of us have
recently studied
\citep{Benhaiem+SylosLabini_2015,Benhaiem+SylosLabini_2017} the
evolution from configurations that are initially ellipsoidal {or of
  an irregular shape} and found them to give rise to a virialized
central core surrounded by very flattened configurations made by both
weakly bound and ejected particles.

These results, combined, have led us to the idea that, with some
initial rotational motion, it might be possible to generate a spiral
structure from these kind of initial conditions.  Indeed, the large
radial velocities are generated in a small region (of the order of the
minimal size reached) in a very short time (much less than one
dynamical time), and thus the radial distance these particles
subsequently travel once they are outside the core
 can be expected,
given approximate conservation of angular momentum, to be correlated
with the integrated angle they move through.

The paper is organized as follows. In Sect.\ref{simulations} we
present the details of our numerical simulations.  Sect.\ref{results}
is then devoted to a discussion of the three-dimensional and
two-dimensional results of our simulations and their relation with
some key observational results on spiral galaxies. Finally, in
Sect. \ref{discussion} we draw our main conclusions.  In the Appendix
we {detail how we constructed the projected velocity maps from our
simulated mass distributions.}

%%%%%%%%%%%%%%%%%%%%%%%%%%%%%%%%%%%%%%%

\section{Simulations}
\label{simulations}
We have considered a very simple set of initial conditions that
combines the characteristics {described above}: breaking of 
the spherical 
symmetry of the initial mass distribution, a velocity distribution that
 
is sub-virial ({or, more generally, out of equilibrium}), and some 
coherent rotation. More precisely, we consider the
following: $N$ particles distributed randomly, with uniform mean
density, inside an ellipsoidal region, and velocities {which
  correspond to a coherent rigid body-like rotational motion
   about
  the shortest semi-principal axis.}  
 {Although these are ad hoc and 
clearly too idealized to describe a physically realistic situation},
 in the context 
of the theory of galaxy formation
these kinds of  initial
conditions {have often been argued}
to be reasonable  (see, e.g.,
\cite{ELS_1962}). {Nevertheless,
they are very different from those described in current scenarios
for galaxy formation in the context of cold dark-matter-dominated
cosmological models, which are characterixed by hierarchical 
collapse. We note, however, that in  
cosmological scenarios with very suppressed initial fluctuations 
at very small scales (e.g. in models with warm dark matter), a
monolithic collapse from a  quasi-uniform initial state may be
a more reasonable approximation.}  
{In any case, our goal
here is to identify and study a physical mechanism and
its possible observational signatures, and not to provide 
a realistic modeling of great complexity.} 

The parameters we choose to characterize our initial configurations
are then (i) the ratios of the semi-axes of lengths $a_1 \geq a_2 \geq
a_3$: the ellipsoids can  be prolate,  
oblate, or triaxial 
and they are specified by the flatness parameter $\iota=(a_1/a_3)-1$;
(ii) the initial virial ratio {$b_{rot}=2K_{rot}/W_0$}, where
$K_{rot}$ is the kinetic energy of the rotational component of the
motion, which has an angular velocity independent of radius {(i.e.
solid body rotation)} and parallel to the shortest semi-principal axis,
and $W_0$ is the initial gravitational  potential energy
\footnote{{Because the force-smoothing at small scales is  a
      factor of 10 smaller than the initial interparticle distance,
    the difference between $W_0$ (computed using the 
      Newtonian potential)} and the Clausius virial term 
      (computed using the exact forces acting on particles) is
    negligible.}  We have explored a large parameter range in this
family of initial conditions, {extending down to $b_{rot}=-1$
  which, although strictly ``virial,''  is well out of equilibrium for
  the chosen velocity distribution. We follow} the evolution under
self-gravity until a time $t \approx (50 \div 100) \tau_d$ where
$\tau_d$ is the characteristic time scale for their mean-field
evolution defined as
\be
\label{taud} 
\tau_d = \sqrt{ \frac{\pi^2 a_3^3}{8 G M}} \;,
\ee
where $M$ is the initial mass and $G$ is the Newton's constant.  All
simulations\footnote{See
  \cite{ejection_mjbmfsl,syloslabini_2012,Benhaiem+SylosLabini_2015,Benhaiem+SylosLabini_2017}
  for details.} are performed for $N=10^5$ particles, using the {\tt
  gadget-2} code \citep{gadget_paper}, adopting {a force-smoothing which is approximately one-tenth of the initial mean
  interparticle separation.}  In this paper we report {in detail}
results for just one chosen simulation, whose features are
representative of this class of models. {Greater details on
  numerical issues and analyses of results for a broad representative
  range of these initial conditions, and also for a range of particle
  numbers extending up an order of magnitude larger,} will be provided
in a separate paper {\citep{benhaiem+joyce+syloslabini_2017}.}

\section{Results} 
\label{results}

\subsection{Three-dimensional properties} 

\begin{figure}
\includegraphics[width=9.5cm,height=9.5cm]{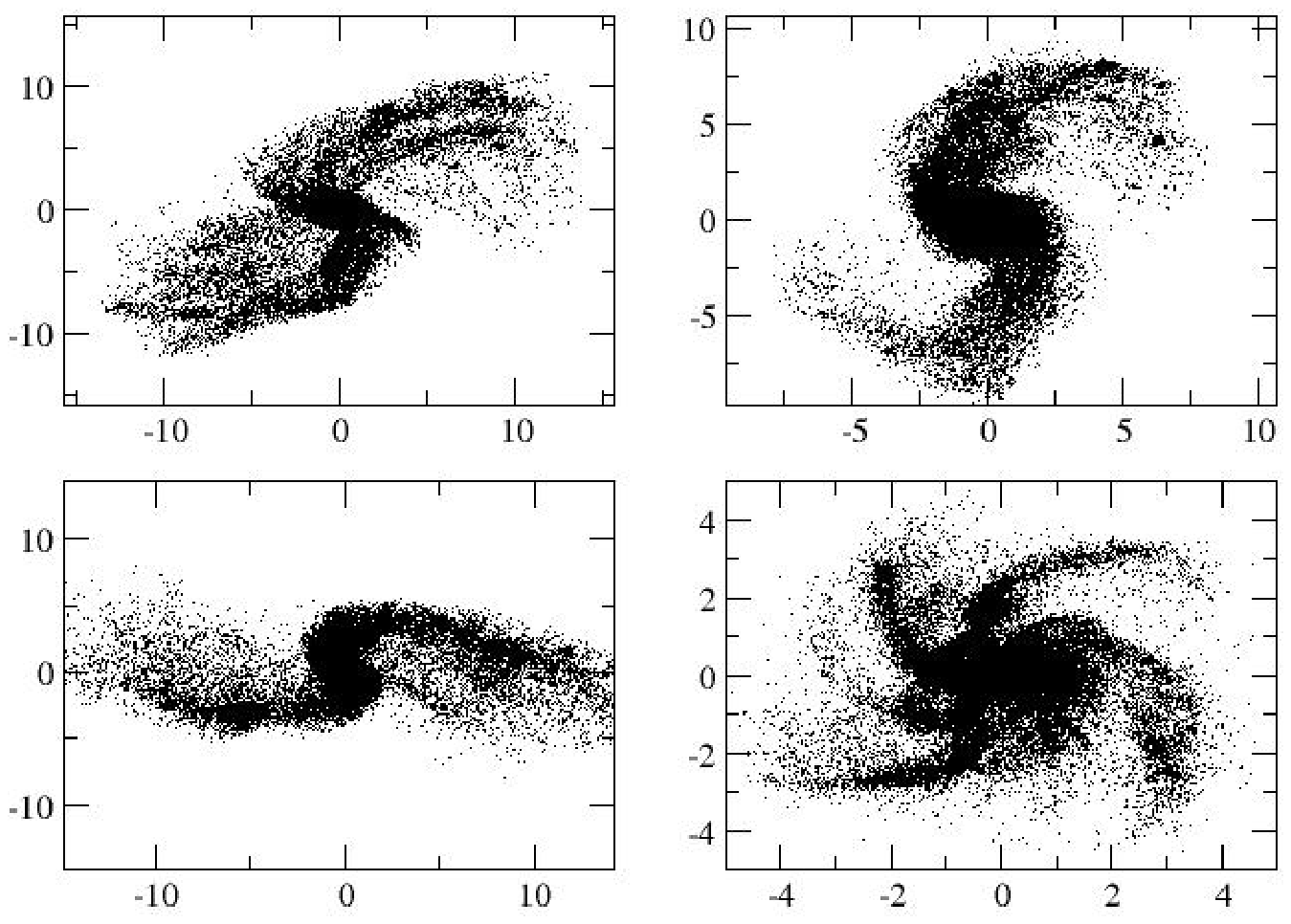}%{Figure0.eps}
\caption{Configurations resulting from four different initial conditions.}
\label{figure0} 
\end{figure}

{We observe, as expected given the chosen initial velocity
distribution and normalization}, a significant contraction and a 
subsequent re-expansion of the system on a time scale $t \sim \tau_d$. 
Associated with this behavior is, as anticipated, also a strong 
injection of energy into a significant fraction of the particles, which are those
initially located furthest from the center (i.e. close to the
semi-major axis) and which pass through the center of the structure
latest during the collapse. Correspondingly, we observe an
amplification of the spatial asymmetry during this phase (with, in
particular, a more rapid contraction along the shortest axis).
In addition to these features, which have been studied extensively in
previous works
\citep{Benhaiem+SylosLabini_2015,Benhaiem+SylosLabini_2017}, we find
 {that these systems are qualitatively characterized in their
outer parts by spiral-like structure}, with a rich variety of forms ---
see Fig.~\ref{figure0} --- ranging from
some qualitatively resembling  more grand design spirals, and others 
qualitatively resembling 
barred
spirals and even flocculent spirals in some cases\footnote{{Here, and 
in the following figures, we use units of length in which $a_3=1$, 
and units of time in which $\tau_d=1$; 
energies are given in units}
in which ${G N m^2}/{a_3}=1$.}. {Detailed analysis of the 
evolving configurations confirms that the emergence of this spatial
organization--- associated with a velocity distribution with very
specific characteristic properties that we will describe below ---
is indeed the result of the injection of energy into some
of the mass around the time of maximal 
contraction, which gives it large radial velocities in
addition to the initial rotational motion.}
 
\begin{figure}
\includegraphics[width=9.5cm,height=9.5cm]{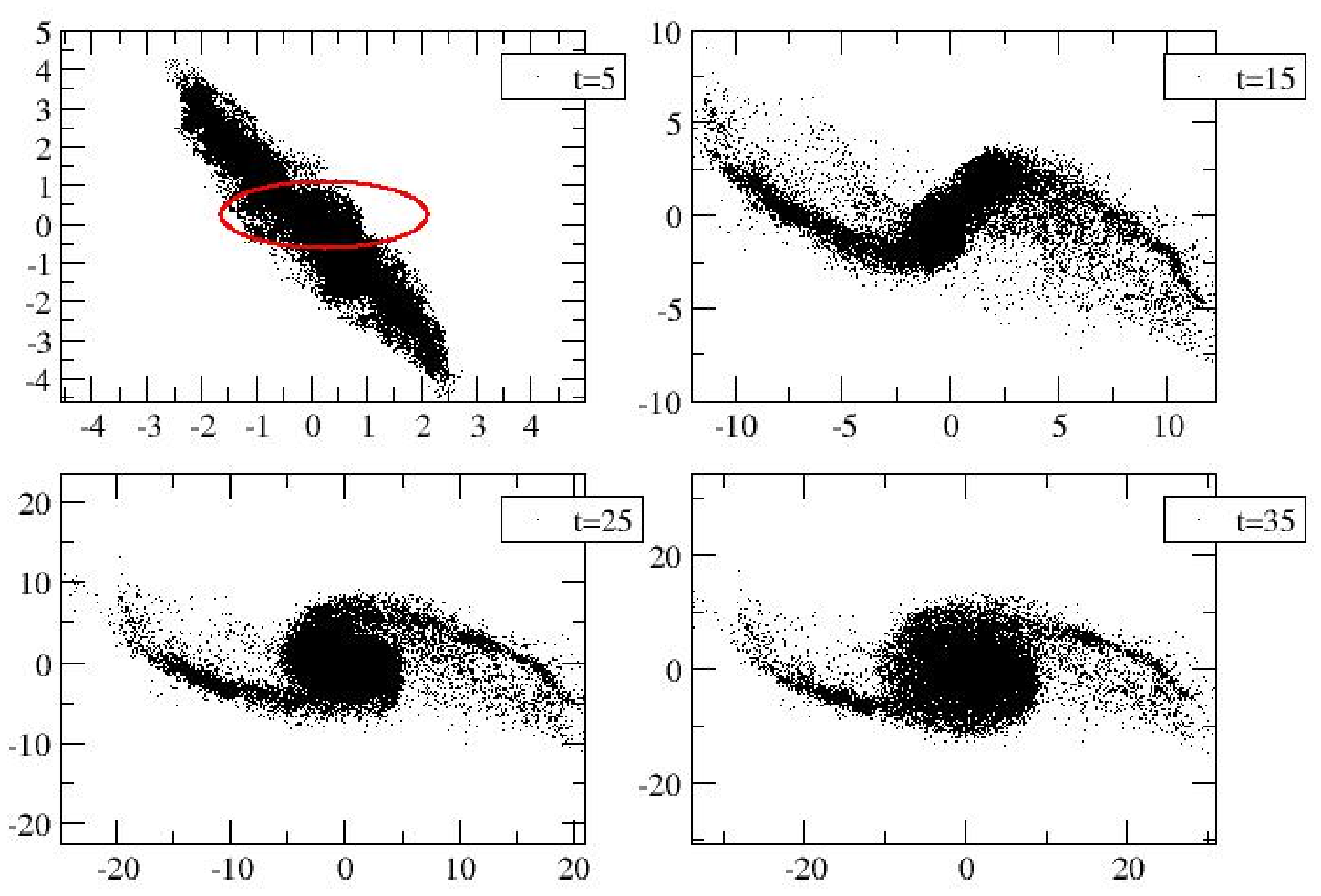}
\caption{Configurations resulting from four different times (see the labels). 
The solid line in the
  upper left panel corresponds to the initial ellipsoid.}
\label{figure1} 
\end{figure}

\begin{figure}
\includegraphics[scale=0.6]{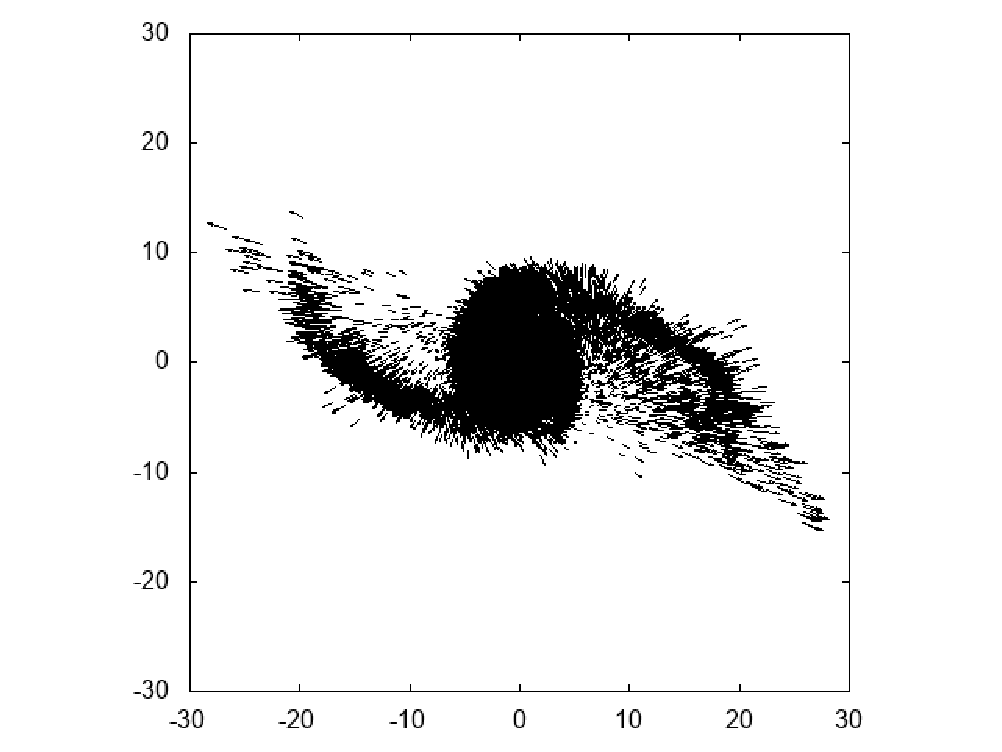} 
\caption{Configuration  at $t=25$: arrows are proportional to velocities}
\label{S9_velplot}
\end{figure}

As {anticipated} above, we focus here, {for simplicity}, on the detailed
analysis of just one specific initial condition,  {with $\iota=1$ 
and $a_2=a_3$ (i.e. a prolate initial ellipsoid), and  $b_{rot}=-1.0$. 
We choose this case because, even if it corresponds to a case 
that is not so far out of equilibrium characterised by a 
less violent contraction and expansion, it produces structure
which is fairly typical of all cases.}  Shown in
Fig.~\ref{figure1} {are configurations of the evolved configuration at
different times \footnote{see {\tt goo.gl/L1fRzZ} for the full movies
of the time evolution.} } projected on the 
plane orthogonal to
the initial 
shortest semi-principal axis, along which the structure is (as
expected) very flattened in extent compared to the observed
projection: diagonalizing the inertia tensor to determine the
principal axes and eigenvalues, we find a typical offset of a couple
of degrees from the initial axes, but a much larger ratio for the
eigenvalues, corresponding to a flatness parameter $\iota \approx 3$,
while the core is triaxial with a flatness parameter $\iota \approx 1$ and
corresponds to a triaxial ellipsoid.
We note that, once formed, the spiral-like arms expand radially,
slowly changing shape.
Indeed the velocity field of the particles in the outer part of
the object is almost radial and directed outward (see
Fig.~\ref{S9_velplot}).

\begin{figure}
  \includegraphics[width=8cm,height=8cm]{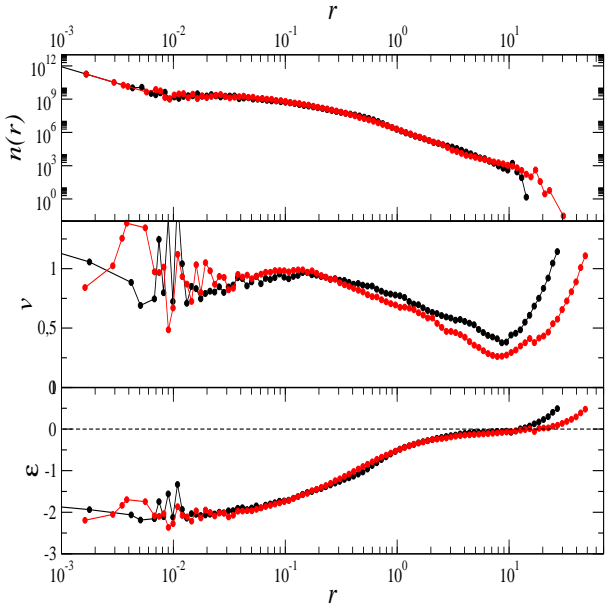}
  \caption{From top to bottom: (i) density profile, (ii) 
    velocity profile, and  (iii)  energy profile, at two
    different times: $t=25$ (black dots) and at $t=45$ (red dots).
    }
  \label{S9_figure4} 
\end{figure}

Fig.~\ref{S9_figure4} shows the density profile $n(r)$, the velocity
profile $v(r)$ and the energy profile $\epsilon(r)$ computed as
averages in radial bins of constant logarithmic width.  During the
time evolution, the outer tail of $n(r)$ is stretched to larger and
larger distances. In general, when the system contraction during the
collapse is strong enough to produce a large change of the particle
energy distribution, the tail of the density profile is well fit by
a power-law behavior with $n(r) \sim r^{-4}$
\citep{syloslabini_2013}. Correspondingly the velocity  and the
energy profiles also extend to larger and larger scales. 
 {At the largest radii, as indicated by the average
value $\epsilon(r)$ particles are unbound (with $\epsilon >0$),
while in the core region particles are strongly  bound (i.e. $\epsilon$ 
below $-1.5$);  there is then an extended intermediate 
region in which many  particles are marginally  bound 
(i.e $0 > \epsilon > -0.5$}).

\begin{figure}
  \includegraphics[width=8cm,height=7cm]{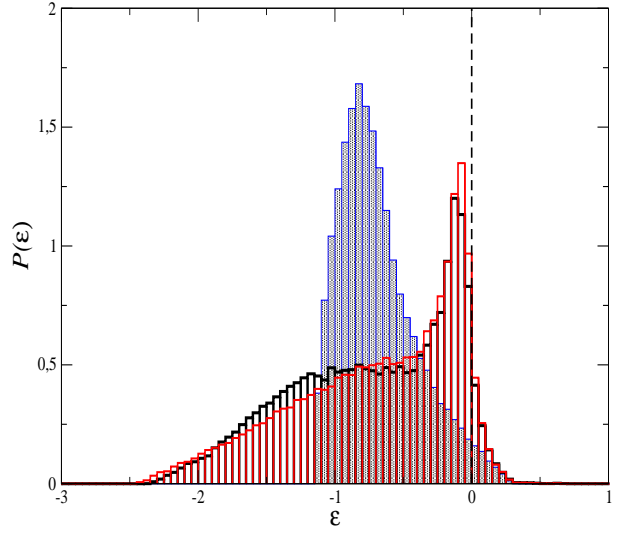}
  \caption{Energy distribution at $t=25$ (black), at $t=50$ (red), and
    at $t=0$ { (green)}. }
  \label{S9_figure5} 
\end{figure}

The energy distribution $P(\epsilon)$ at two different times
($t=25, 50$) together with that of the initial conditions, is shown in
Fig.~\ref{S9_figure5}: we note that a large change of
$P(\epsilon)$ has occurred during the gravitational collapse of the
cloud at $\approx \tau_d$ while at later times the shape of the
distribution remains approximately the same. %

\begin{figure}
  \includegraphics[width=8cm,height=7cm]{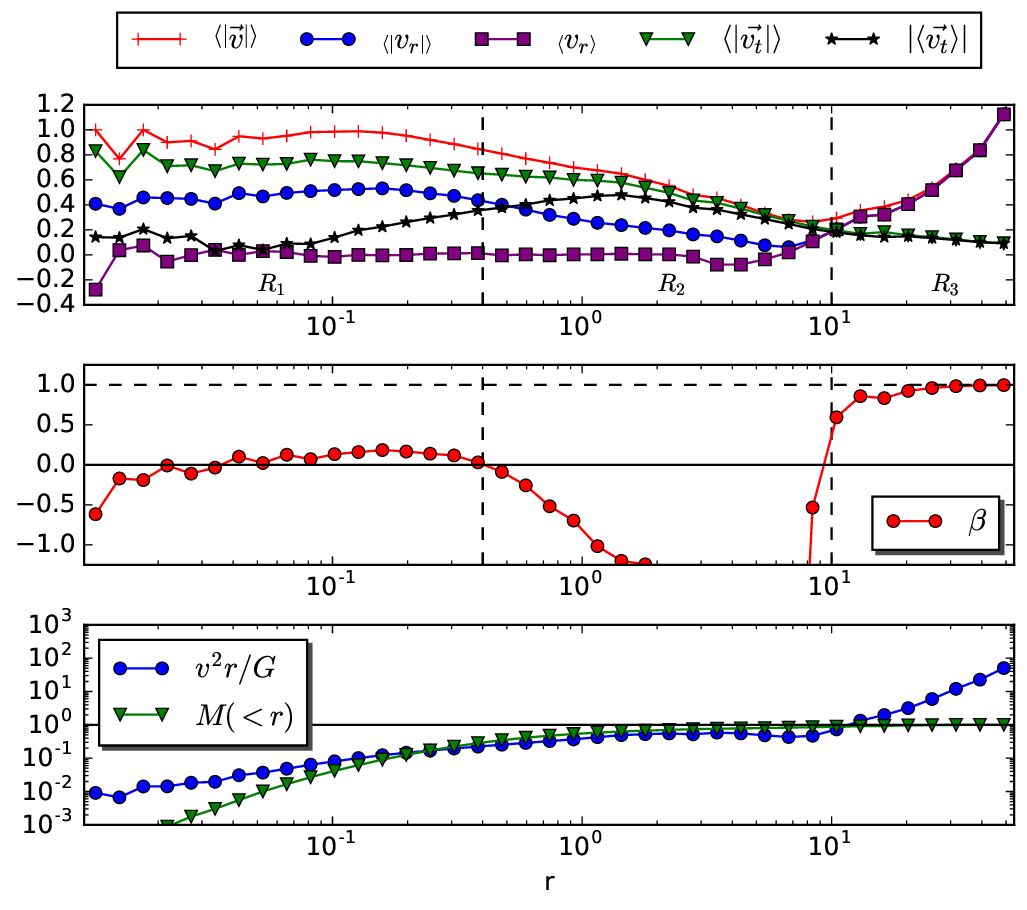}
        \caption{Configuration at $t=45$. Upper panel: components of
          particle velocities averaged in spherical shells as a
          function of radius.  Middle panel: anisotropy parameter
          $\beta(r)$.  Lower panel: mass estimated from the velocity
          assuming stationary circular orbits, and the actual enclosed
          mass.}
  \label{figure2}
\end{figure}

The upper panel of Fig.~\ref{figure2} shows the average, in spherical
shells of radius $r$, denoted $\langle \cdots \rangle$, of the radial
{component
\[
{v}_r =
\frac{\vec{r} \cdot \vec{v}}{{r}}
\] 
of the velocity, and of the ``transverse" velocity
\[\vec{v}_t = \frac{\vec{r} \times
\vec{v}}{{r}}\;,
\]
defined parallel to the angular momentum relative to the origin
(at the center of the structure).}
Thus, in particular, a coherent rotation of the shell
in a plane corresponds to $\langle | \vec{v}_t| \rangle$= $|\langle
\vec{v}_t \rangle|$.

The middle panel of Fig.~\ref{figure2} shows the
velocity anisotropy
\[\beta (r) = 1 - \frac{\langle v_t^2 \rangle}{2
  \langle v_r^2 \rangle} \;.
  \]
  Finally, the lower panel of Fig.~\ref{figure2} shows $v^2 r/G$, the
  mass that would be enclosed inside this radius if the motions were
  purely circular and the mass distribution spherically symmetric, and
  the mass $M(<r)$ actually enclosed inside the radius $r$.

According to the behaviors observed, we can divide the structure into
three regions: (i) an inner part (R1) in which, as
\[
\langle |\vec{v}_t
| \rangle \gg |\langle \vec{v}_t \rangle| \,,
\] 
there is no significant
net rotation, and given that $\beta \approx 0$, the velocity
distribution is close to isotropic;
(ii) an intermediate range of radii (R2), extending over about a
decade, in which $\beta$ deviates strongly from zero as a net coherent
rotational motion develops and dominates at larger radii, i.e.
\[ \langle |\vec{v}_t | \rangle \approx |\langle \vec{v}_t
\rangle | \gg \langle |{v}_r| \rangle \;; 
\] 
correspondingly (lower panel of Fig.~\ref{figure2}), there is a good
agreement between the estimated and actual enclosed mass
{in this region};
and (iii) an outer region (R3) in
which the rotational motion of the particles is still coherent, but
radial motions, {with almost negligible dispersion, are now
  predominant, i.e.
  \[
  \langle |\vec{v}| \rangle \approx \langle |{v}_r|
  \rangle \approx |\langle {v}_r \rangle |
  \;.
\] 
Region R3 is also characterized clearly by the behavior of the
estimated enclosed mass, which greatly overestimates the actual
enclosed mass. This reflects the fact that the mass is weakly bound or
even unbound rather than bound on circular orbits.

Measurement of the particle energies (see Fig.~\ref{S9_figure4}) shows
that the transition from R2 to R3 is indeed approximately that from
unbound to bound particles, and that in the outer part of R3 all
particles are unbound. Indeed, asymptotically the {knee} between the
two regions is precisely the transition from bound to unbound orbits,
with the shell with {$\langle |v_r| \rangle \approx 0$} corresponding
to particles with zero energy. At large distance in R3 we have,
correspondingly, a linear growth with distance of the radial velocity
that is simply a reflection of the ballistic radial motion. Thus,
when we study these curves as a function of time, region R1 and the
inner part of R2 are stationary to a very good approximation, while
the boundary with R3 propagates out progressively and the size of R3
itself grows linearly in time, with the maximal velocity remaining
fixed.

These angle-averaged data do not give information about the angular
dependence of the radial velocities in particular, which are very
non-trivial: the presence of the spiral-like structure visible in
Fig.~\ref{figure1} is a reflection of the fact that the particles'
transverse motions become correlated with their radial velocities
because of the approximate conservation of angular momentum (and
energy) of the ejected particles, which, once outside the core, move
in an approximately stationary central potential.
The particles forming the spiral structure  preferentially  have a
radial velocity oriented along directions close to the initial
longest semi-principal axis, 
and the structure is elongated the most along 
directions in which the radial velocities are maximal.  
Clearly, the
precise form of the spiral structure depends directly on the
dispersion of the energies of the high-energy particles at the time of
collapse compared to their transverse velocity at this time {(and
  thus, in particular, on the parameter $b_{rot}$)}.  The
non-stationary nature of the structure also 
manifests itself in the evolution  of the form of the spiral structure. 
In particular, it becomes
more elongated (and less axisymmetric) in time.

\subsection{Estimation of Typical Length/Time Scales}

{Even if our models are too simple
and idealized to be meaningfully confronted with
observations in any great detail, we can consider 
the {\it qualitative} compatibility of the features 
of the mass distributions generated with the 
observed properties of real astrophysical
systems. In particular, we focus here on the
primary astrophysical motivation for our study
--- spiral galaxies --- although, as noted, 
several other applications could also be
explored}. For any such comparison we 
evidently  need to relate 
approximately relate the scales of our toy model to 
physical scales. Bearing in mind that the typical scale
of observed rotation velocities in disk galaxies 
is $200$ km/sec, we define the dimensionless
parameters  as follows: 
\[
v_{200}= \frac{a_3/\tau_d}{200} \;
 \mbox{km/sec} \;.
\] 
We can then write
\[
a_3   \approx \left( \frac{200 \, v_{200}}{n}
\times  t_{\rm Gyr}\right)  \,{\rm kpc}
\]
where $n$ is the number of dynamical times in our 
simulation and $t_{\rm Gyr}$ is its duration given in billion 
of years.  Thus, for $n \approx 50$ as in Fig.~\ref{figure1}, 
and taking $t_{\rm Gyr} \sim 1$, which corresponds
to a mass (by using Eq.\ref{taud})
\[
M = \frac{\pi^2 a_3^3}{8 G \tau_d} \approx 10^{11} M_{\odot} \;, 
\] 
we have that region $R_1$ extends to $\sim 2 \,{\rm kpc}$, and region
$R_2$ extends to $\sim 50 \,{\rm kpc}$. Thus, in order to have a structure
that would possibly be compatible with the typical size of spiral
galaxies, we need to assume that the collapse process which generated
the disk and arms occurred much more recently than the formation of
the oldest stars in these galaxies (with an age $\sim 10$ Gyr). This
is very different from the usual hypothesis that the disk, and its spiral
structure, are at least as old as the oldest stars. From the
observational point of view, however, there is no definitive evidence
establishing the age of spiral arms; rather several observational
studies have suggested that spiral arms are not long-lived
\citep{Elmegreen_etal_1989,Vogel_etal_1993,Tully_Verheijen_1997,
  Henry_etal_2003}.

\subsection{Characteristic of spiral arms} 

 {We note that the arms formed in our models are always trailing.
   This is a simple consequence of the approximate conservation of
   angular momentum for the outgoing particles, which means that the
   transverse components of their velocities decrease with their
   radial distance.  Although, as mentioned, a rich variety of forms
   of the arms can be obtained with different initial conditions, two
   dominant arms as in our chosen simulation are very easily produced,
   with pitch angles of the order of tens of degrees. Thus, our model naturally
   reproduces  very common features of spiral galaxies, which
   are very difficult to explain within the much explored framework of
   density wave theory \citep{Dobbs_Baba_2014}, although density
   variations associated with spiral arms in our models are larger
   than they are in reality.

\subsection{Apparent velocity maps}

Let us now consider  further the compatibility of large-scale motions
of our generated mass distributions with observed apparent motions
{in disk galaxies}.
Depending on the initial conditions we choose, the details of the
kinematic properties will change (e.g. the exact radial dependence of
the velocities), but it is a generic property of this
mechanism of generation of the spiral structure that there is a clear
transition from predominantly rotational motion to predominantly
radial motion, the latter being in the outermost parts the ballistic
motion of freed particles. {This is the case simply because 
the particles that are furthest from the center at long times
are unbound or very loosely bound outgoing particles that have 
lost almost all their transverse velocity because of angular 
momentum conservation}.  Let us focus on this characteristic
feature.

Decades of study of various different observational tracers of the
velocity fields provide strong evidence for predominantly rotational
motions in disk galaxies \citep{Sofue_Rubin_2001}.  {For what concerns
  our Galaxy, in which apparent motions have been measured over four
  decades in scale \citep{Sofue_2017}, the angular dependence of the
  projected velocities, inferred from HI emission in particular, shows
  convincingly that the motion of the disk is very predominantly
  rotational up to a scale of order $15$ kpc
  \citep{Kalberla_Dedes_2008, Sofue_2017}: as mentioned above, such a
  coherent rotation is also characteristic of the region R2 in our
  models. For this reason, the key observation thus concerns the
  nature of the motion at large distance, i.e. $> 15$ kpc, in our
  Galaxy.  In this respect it is interesting to note that there is
  nevertheless also evidence for significant coherent radial motions
  beyond a few kpc and increasing with radius \citep{MLC_CGF_2016}.
  Beyond this scale the constraints are much weaker, but in the near
  future measurements from the GAIA satellite \citep{Gaia_2016} will
  make it possible to distinguish the nature of the motions at much
  larger scales.  In addition the GAIA satellite will be able to shed
  light on the nature of hyper-velocity stars that are unbound from
  the Milky Way and shows a surprising anisotropic distribution
  \citep{Brown_2015}.  A population of such stars might
  {possibly} correspond to ejected particles in our models.

{Let us now consider  constraints on velocity fields from 
external disk galaxies.
In this case apparent (LOS) velocities are probed robustly out to
scales of several tens of kiloparsecs, and in some cases even larger
\citep{Sofue_2017}, but the strength of evidence for rotation 
depends on the scale and weakens at larger scales.}
These measurements are both one-dimensional (i.e.,
along the major axis of the observed galaxy) and two-dimensional
(mapping out the full projected velocity field). The former
measurements provide a direct measure of rotational velocities, but
only on the assumption that the galaxy is in fact a rotating disk: in
this case the major axis of the projection (which is an ellipse)
is orthogonal to the LOS, and thus motions parallel to the LOS must be
rotational. In our models, the mass distribution is not a disk---
indeed it is clearly non-axisymmetric at larger radii--- and,
furthermore, as we have noted, there is intrinsically a strong correlation
of the direction of the outer radial velocities with the intrinsic
longest semi-principal axis.

As a result, we show below that there is generically a
contribution, which may be very large, along the projected 
major axis. {In our models, as a result, even at 
length  scales where the motion is purely radial, we will 
infer a non-trivial rotation curve from a one-dimensional
measurement.}

{For two-dimensional data} the evidence for predominant 
rotation (and the inferred rotation curves) is based on the quality 
of best fits to rotating axisymmetric disk models provided by 
two-dimensional data.
In particular, two-dimensional velocity maps of numerous galaxies
show the pattern {distinctive of} a rotating disk: the alignment of
the kinematic axis --- that along which there is maximal variation of
the projected velocities --- with the projected semi-major axis. Such
alignment is, however, far from perfect and very significant angular
offsets are frequently observed (and attributed to the breaking of
axisymmetry by bars). Furthermore, very significant residuals are typically
measured in such fitting procedures--- typically of the order of $30\%$ or
even larger--- and these are attributed to radial motions (see
e.g. \cite{Erroz-Ferrer_etal_2015}).

To  {qualitatively} evaluate whether the radial motions 
that are dominant in the outer parts of the spiral structure 
in our models can be strongly excluded by observations, as 
one might naively expect, we have thus examined
whether the projected motions {of our toy galaxies}
can provide fits to rotating disk models of a comparable quality 
to those provided by the observed galaxies. To do so, as  {detailed} in 
the Appendix,  using our distributions we have generated  the
projected LOS velocity maps $v_{los}$ of random observers,
characterised by two angles: the inclination angle $i$, defined as the
angle between the vector $\vec{u}_o$ giving the orientation of the
observer's LOS and the vector $\vec{u}_g$ in the direction of the
shortest semi-principal axis 
of the model galaxy, and an angle $j$ defined as the
angle between the projection of the LOS in the plane of the galaxy and
its longest semi-principal axis.  To fit the resulting two-dimensional map with a
rotating disk model, for we determine the velocity as a function of
distance along the axis of maximal variation, and use it as the input
rotational velocity for a rotating disk, for which we analytically determine
 the projection.
Shown in Fig.~\ref{Fig-S9-i30j0b} are the projected velocity maps for
the same simulation analyzed above, for an observer with $i=30^\circ$
and $j=30^\circ$. The maps have been averaged on a grid of size $64^2$
(mimicking the finite resolution of measured maps); the different
panels show the following:
\begin{itemize}
\item
  (i) the two-dimensional  projection of the mass distribution, with the 
  kinematic axis and the major axis {of the projection}
  indicated; in this case the angle between the two axes is about $40^\circ$;
\item
  (ii) the two-dimensional  LOS velocity dispersion map; the largest dispersion
 is in the core where the velocities are isotropic;
\item
  (iii) the two-dimensional  LOS velocity map;
\item
  (iv) the two-dimensional  LOS  velocity map in which the radial velocities
have been removed, illustrating that the motions are indeed very
predominantly radial; 
\item
  (v) the two-dimensional LOS velocity map of the best-fit
  rotating disk
   model (this is obtained by using the one-dimensional LOS velocity profile 
  along the estimated kinematic axis);
\item (vi) the two-dimensional  LOS  velocity residual map.
\end{itemize} 

\begin{figure*} 
\subfloat{\includegraphics[width = 2.5in]{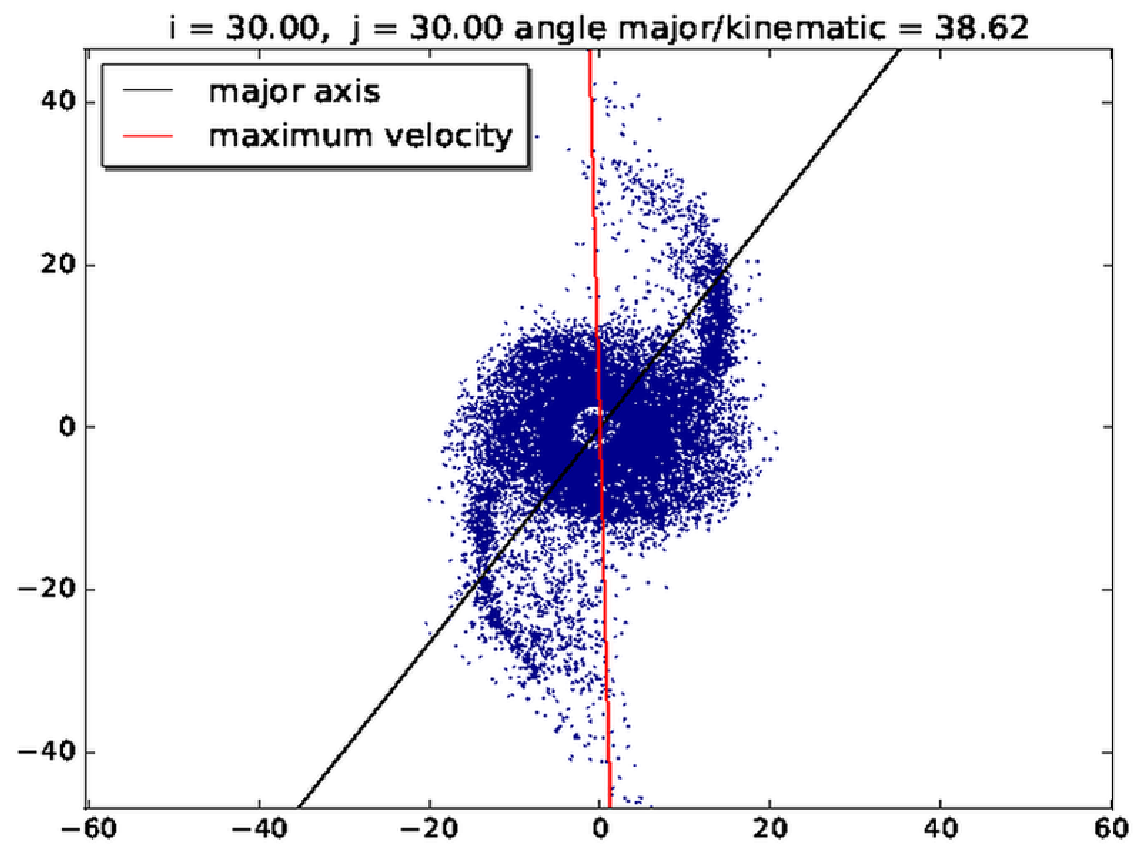}}  
\subfloat{\includegraphics[width = 2.5in]{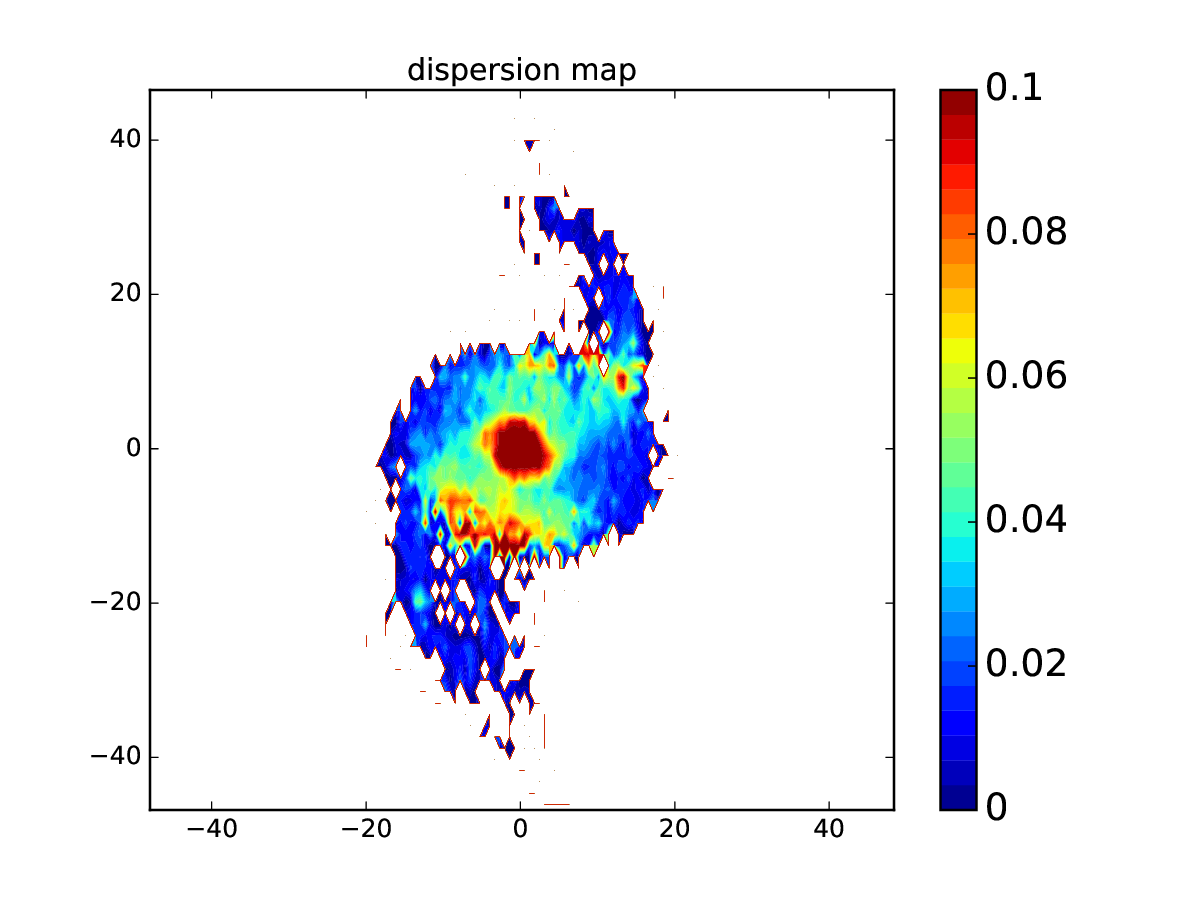}}
\subfloat{\includegraphics[width = 2.5in]{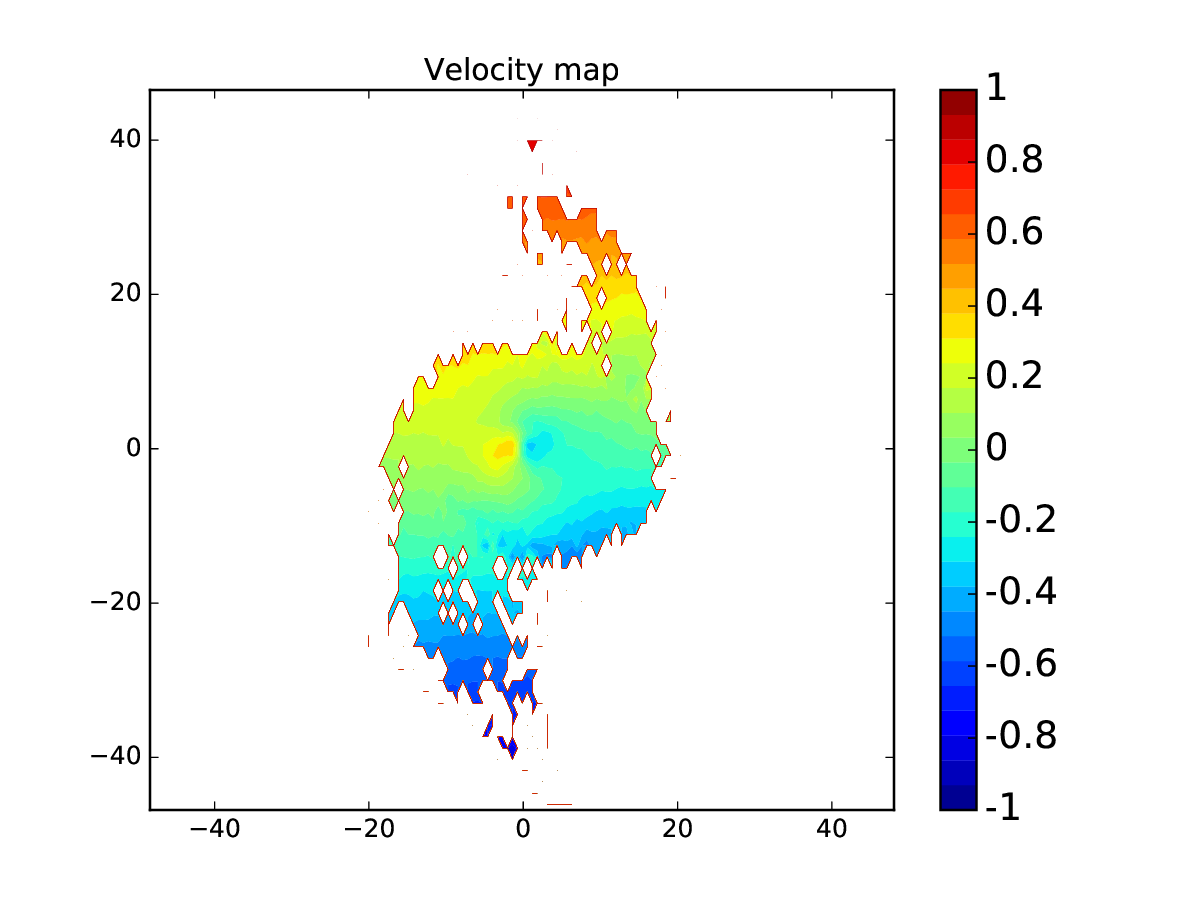}}\\ 
\subfloat{\includegraphics[width = 2.5in]{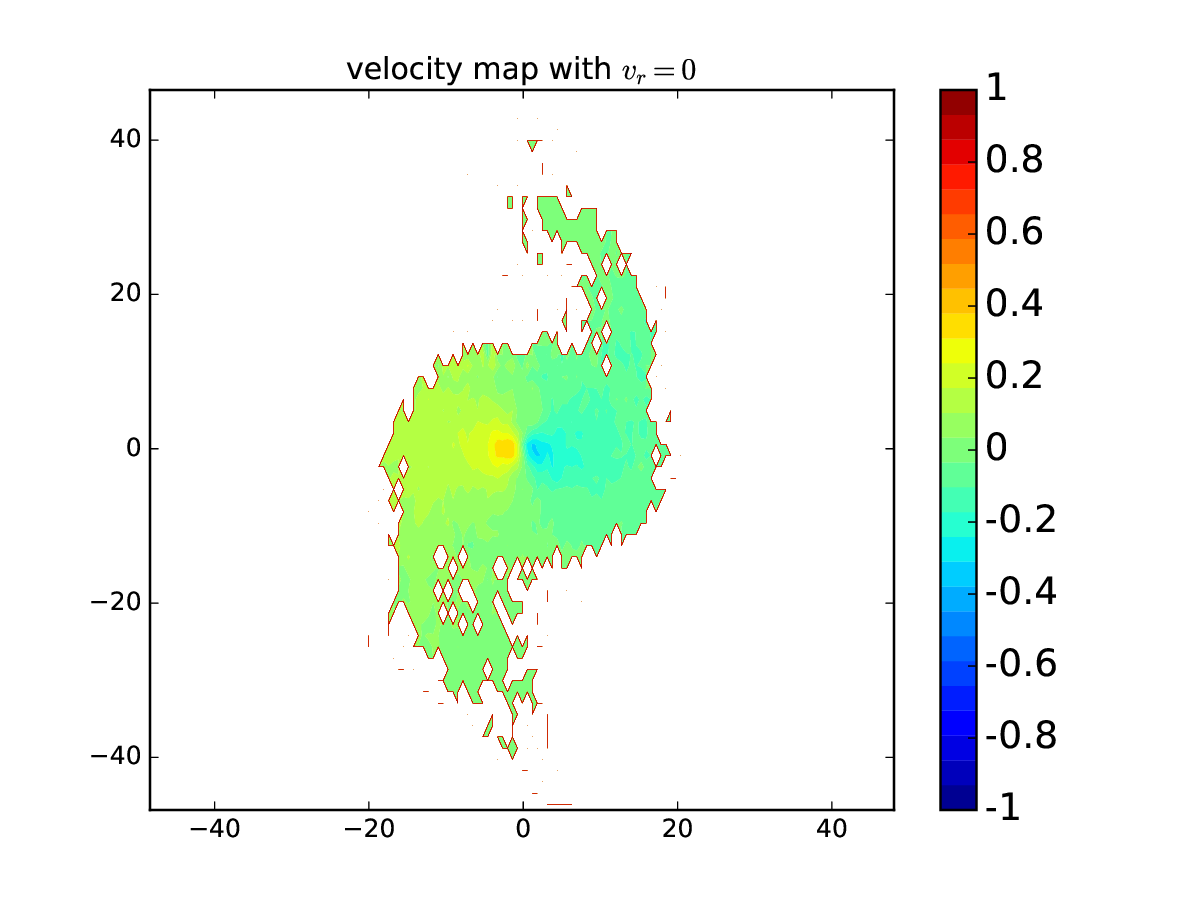}} 
\subfloat{\includegraphics[width = 2.5in]{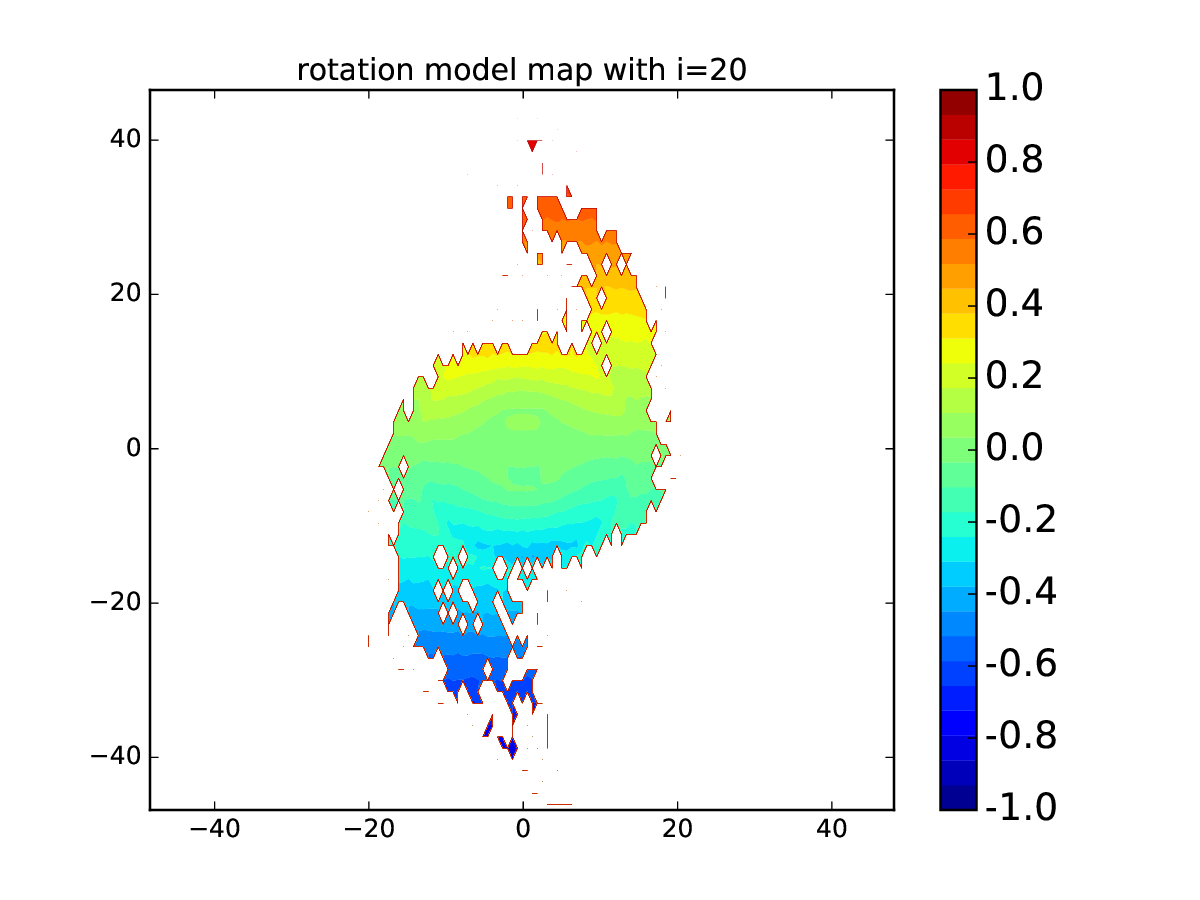}}
\subfloat{\includegraphics[width = 2.5in]{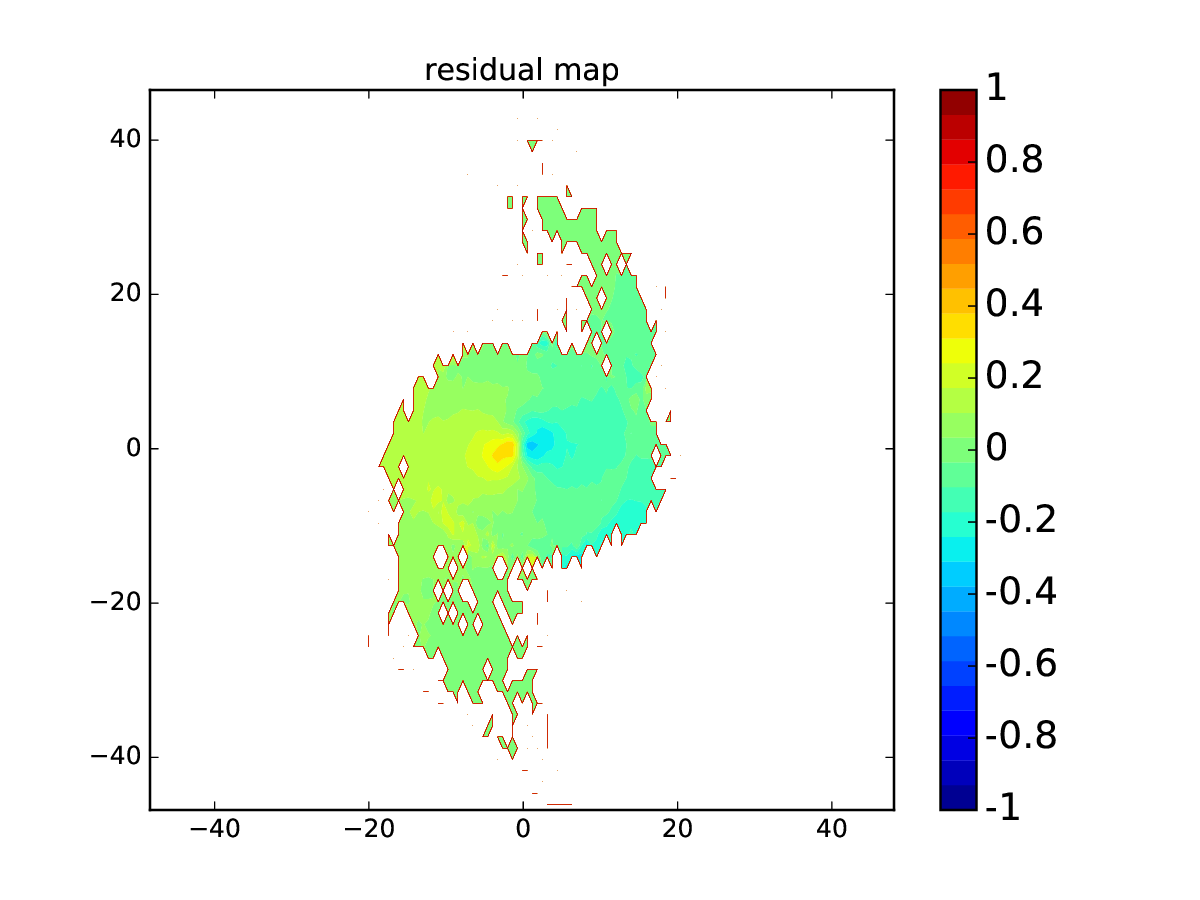}} 
\caption{Projection for  $i=30^\circ, j=30^\circ$ (from top to bottom).
    Left panel: projection of the object on the observer's sky; the
    kinematic axis (red) and the major axis (black) are shown.
    Middle panel: two-dimensional LOS velocity dispersion map.
    Right panel: two-dimensional LOS velocity map.  
    Left  panel:
    two-dimensional LOS velocity for the case in which the 
    three-dimensional radial
    velocity has been set to zero.
    Middle panel: two-dimensional rotational map derived from
    the LOS velocity profile.  
    Right panel: two-dimensional residual
    map.
     }
  \label{Fig-S9-i30j0b}
\end{figure*}

Figure~\ref{Fig-S9-i30j0a} shows, respectively, 
the one-dimensional LOS velocity profile along the kinematic axis and along the axis
orthogonal to it (upper panel) and (lower panel)  the mass estimated by
assuming that the velocities are circular, and the actual mass (i.e.,
by using Eq.\ref{mcr}).
\begin{figure} 
  \includegraphics[scale=0.45]{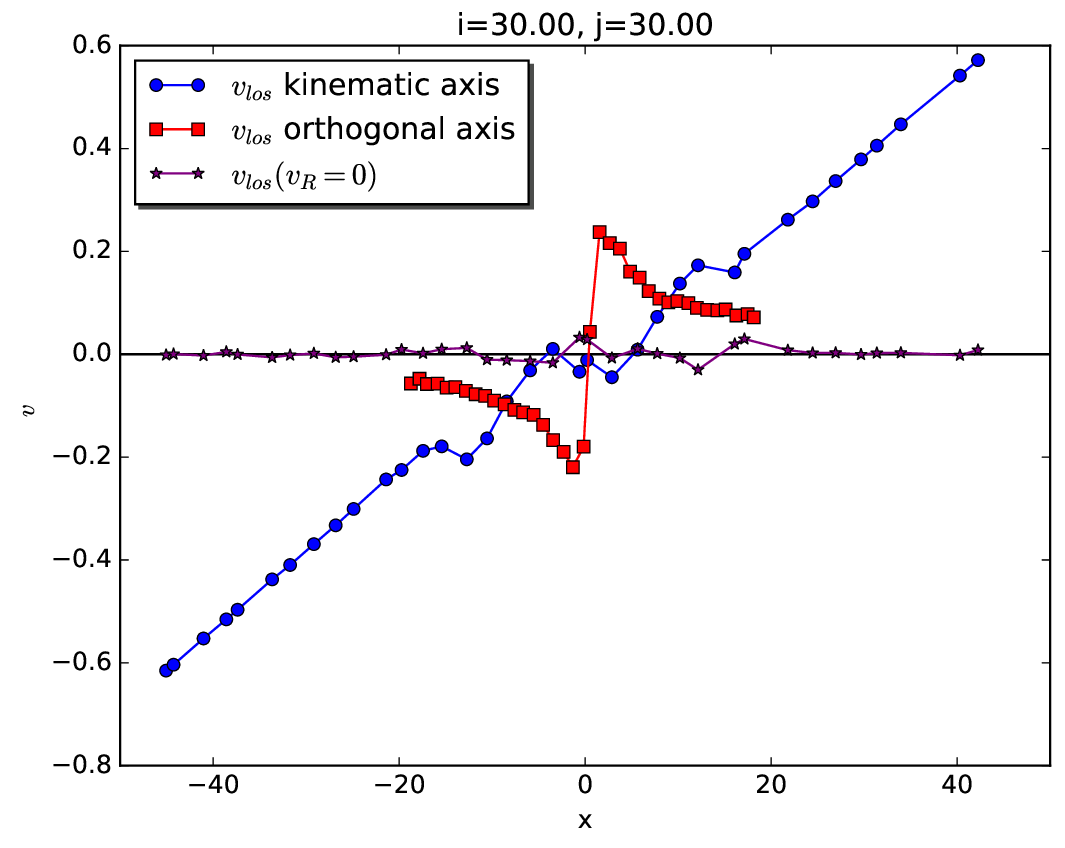}
  \includegraphics[scale=0.455]{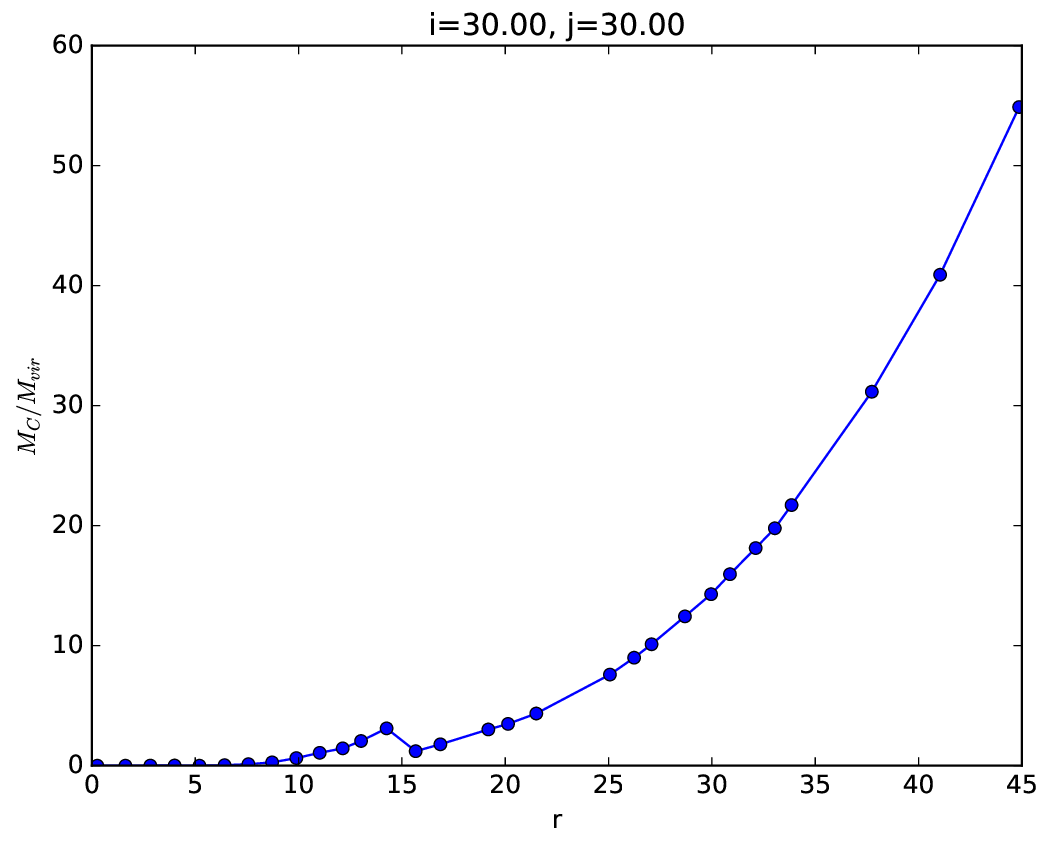}
  \caption{Projection for $i=30^\circ, j=30^\circ$.
    Upper panel: LOS velocity profile along the kinematic axis and
    along the axis perpendicular to it.  Bottom panel: ratio between
    the mass estimated from the LOS velocity (assuming it to be
    circular and stationary) and the actual mass.
           }
  \label{Fig-S9-i30j0a}
\end{figure}

We have then explored (see Fig.~\ref{Fig-S9-i60res}) the full
range of $i$ and $j$. Only for $j$ very close to $\pi/2$ (i.e. an
observer with an LOS almost exactly orthogonal to the axis
along which radial velocities are maximal) do we {fail} 
to obtain a fit to a rotating disk model with residuals 
compatible with the level {reported in the literature
for such fits applied to observational data}. These residuals are
small in all cases, i.e. of the order of $10\%-30\%$ except  for $j
\rightarrow 90$ in which they can be as high as $\approx 50\%-70\%$.
In these images one can discern clearly that  our model galaxies are 
non-axisymmetric at larger radii and, as we have noted, that there is
a strong correlation of the direction of the outer radial velocities with the 
intrinsic major axis: the velocities in the outer parts of the structure are 
radial and very preferentially oriented along the axis which is significantly 
elongated in the structure. As a result there is generically a contribution 
from these radial velocities along the projected major axis.  
In addition, except for very small inclination angles,
the projection of the three-dimensional major axis
is typically very close to the major axis of the projected
image, and the large radial velocities project out their
component along this latter axis. There is thus in practice  
a rough degeneracy between rotating disk models {
with significant, but sub-dominant, radial motions} 
and non-axisymmetric  models with a specific pattern 
of radial velocities {like the one in our models}. 
This is very clearly illustrated by comparing 
in each case the map in which the three-dimensional radial velocity 
is set to zero and the map of the best fit rotational model: despite
the fact that most of the signal at large scales comes from the
radial velocities they can be fit quite well by the rotational
model. 

\begin{figure*} 
\subfloat{\includegraphics[width = 2.in]{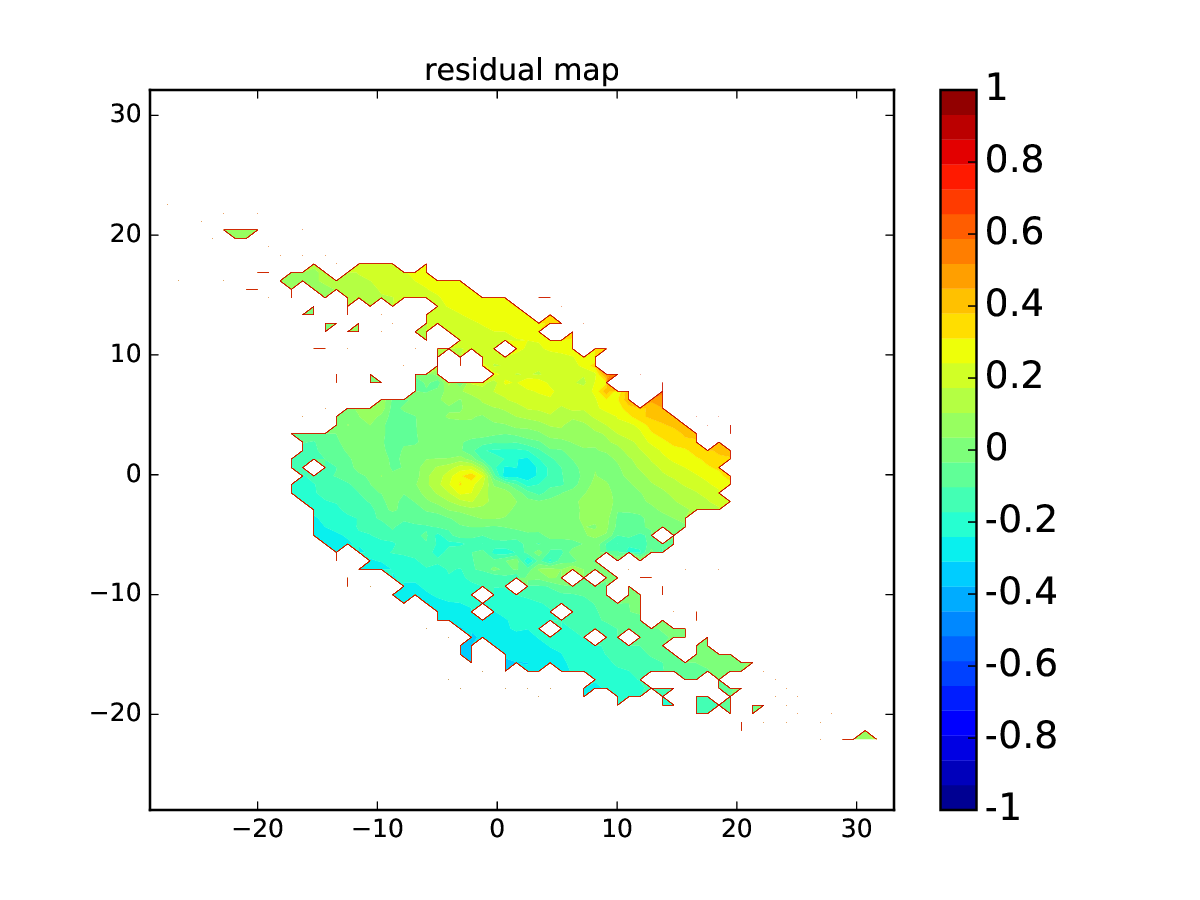}}  
\subfloat{\includegraphics[width = 2.in]{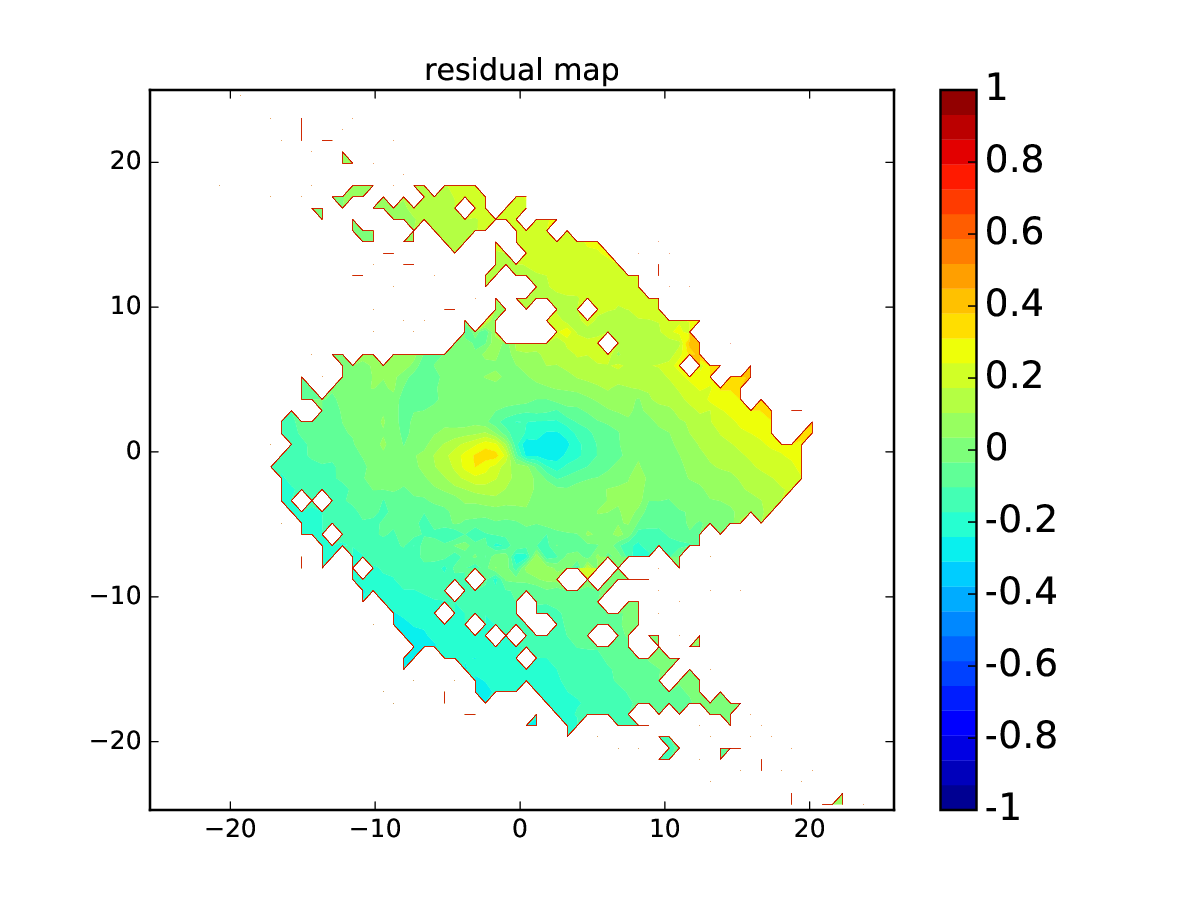}} 
\subfloat{\includegraphics[width = 2.in]{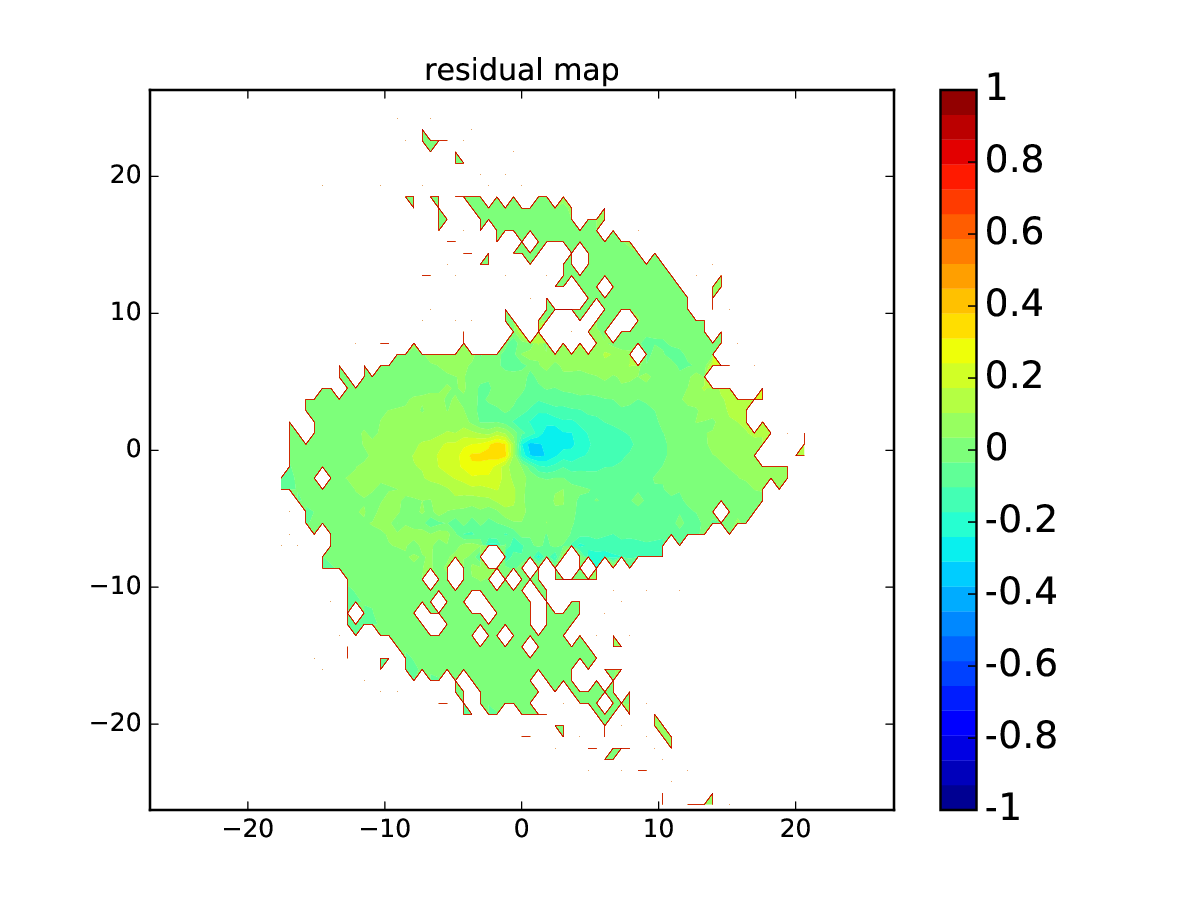}}\\ 
\subfloat{\includegraphics[width = 2.in]{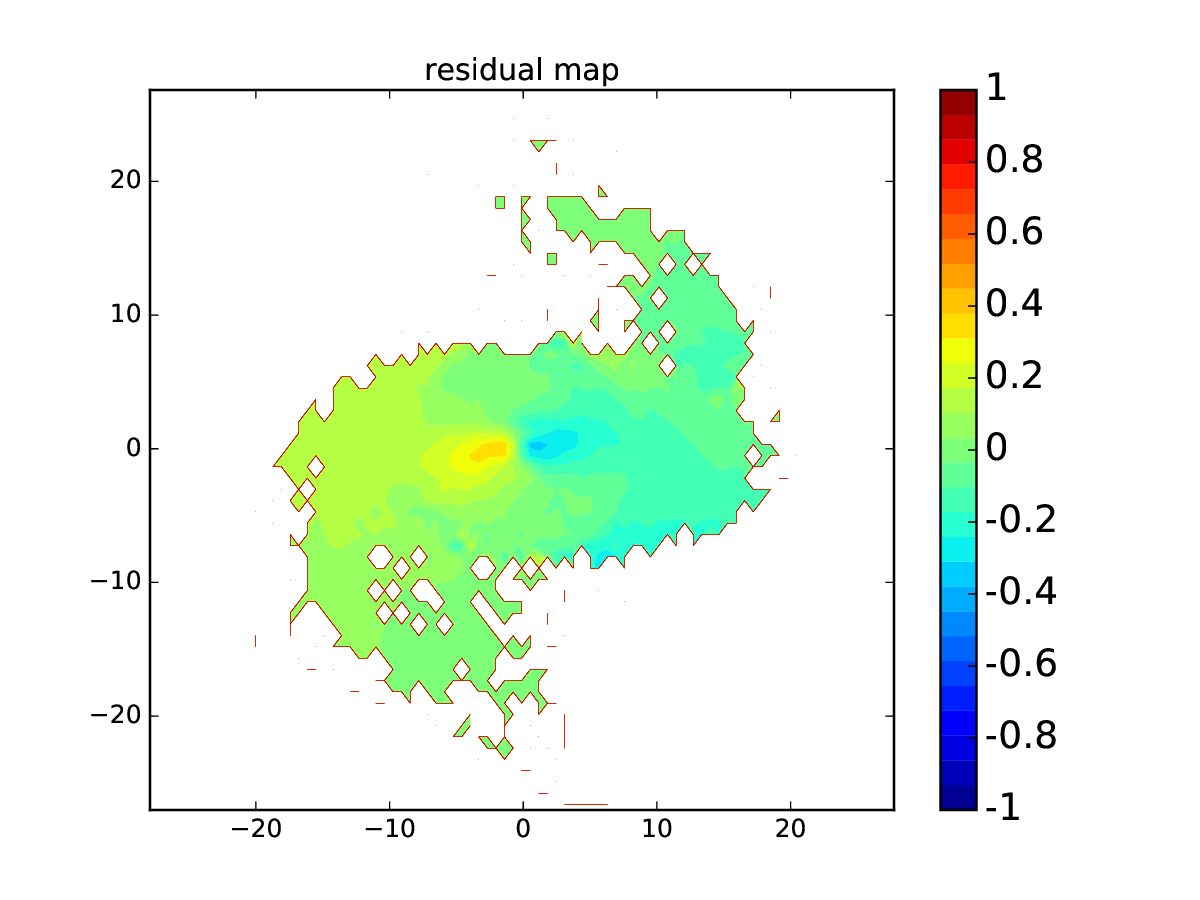}}  
\subfloat{\includegraphics[width = 2.in]{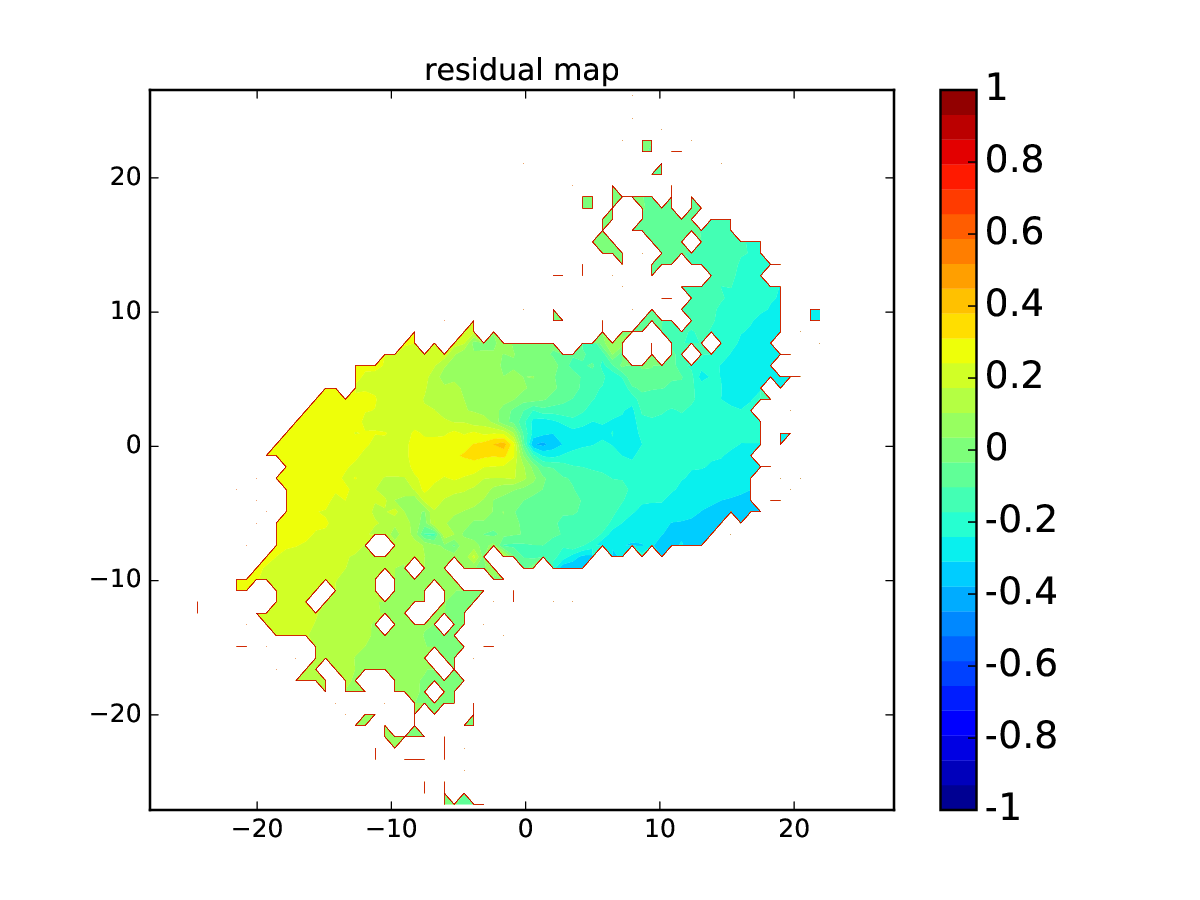}} 
\subfloat{\includegraphics[width = 2.in]{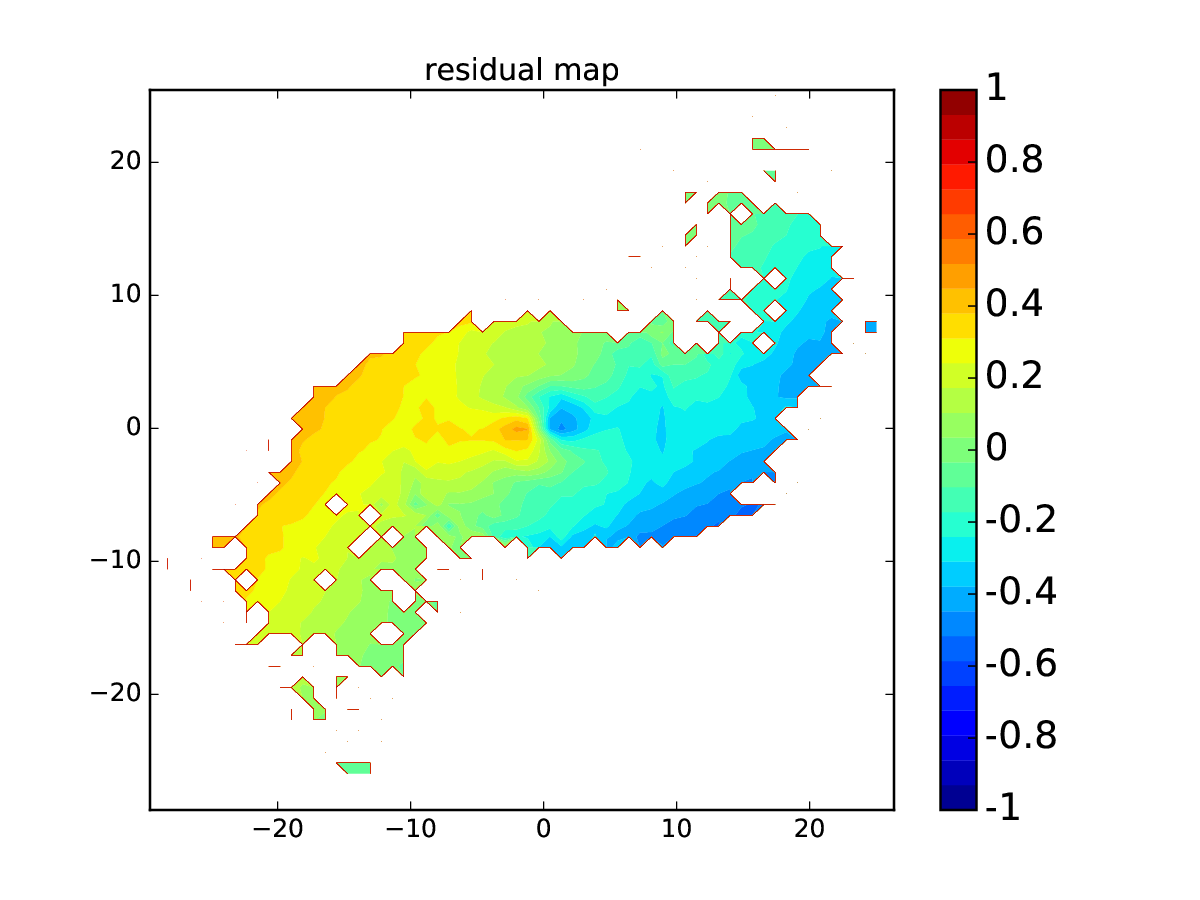}}\\  
\subfloat{\includegraphics[width = 2.in]{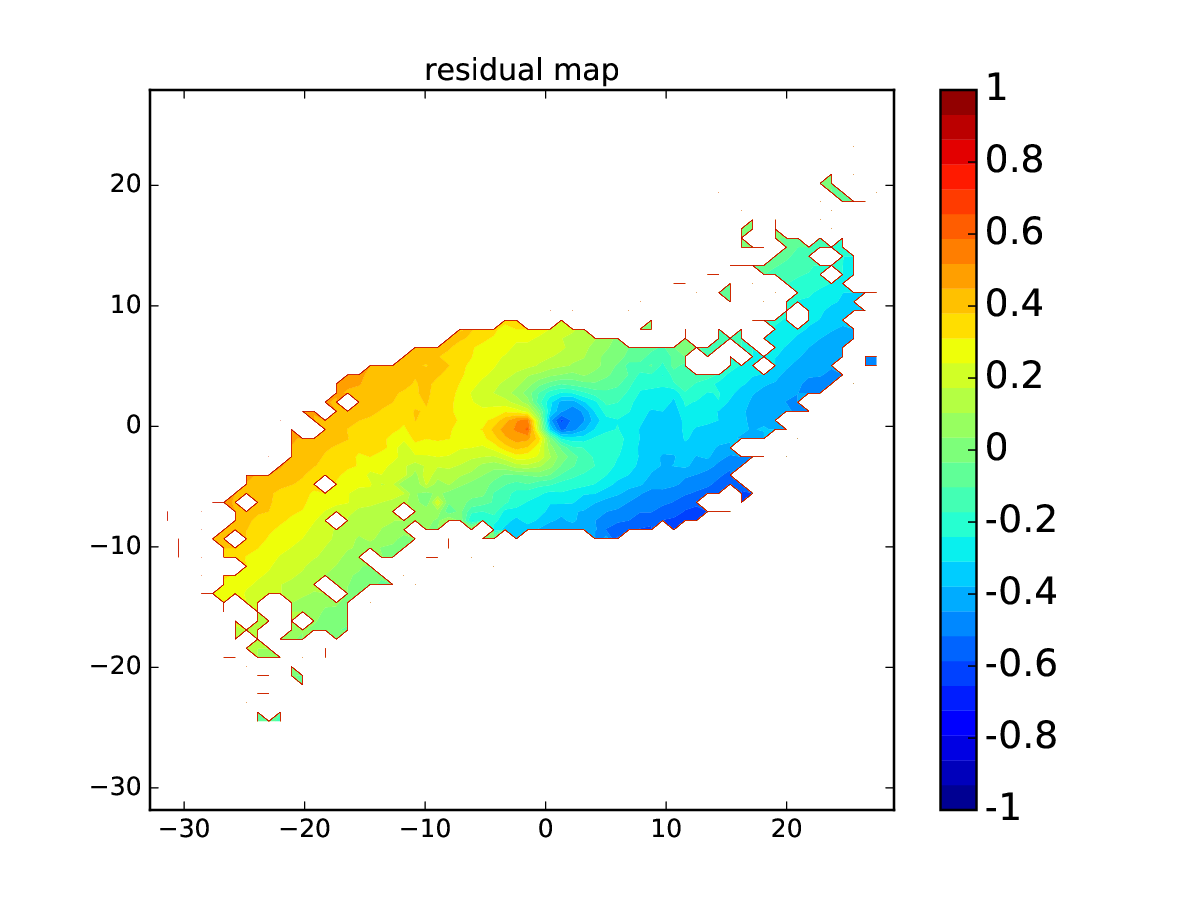}}  
\subfloat{\includegraphics[width = 2.in]{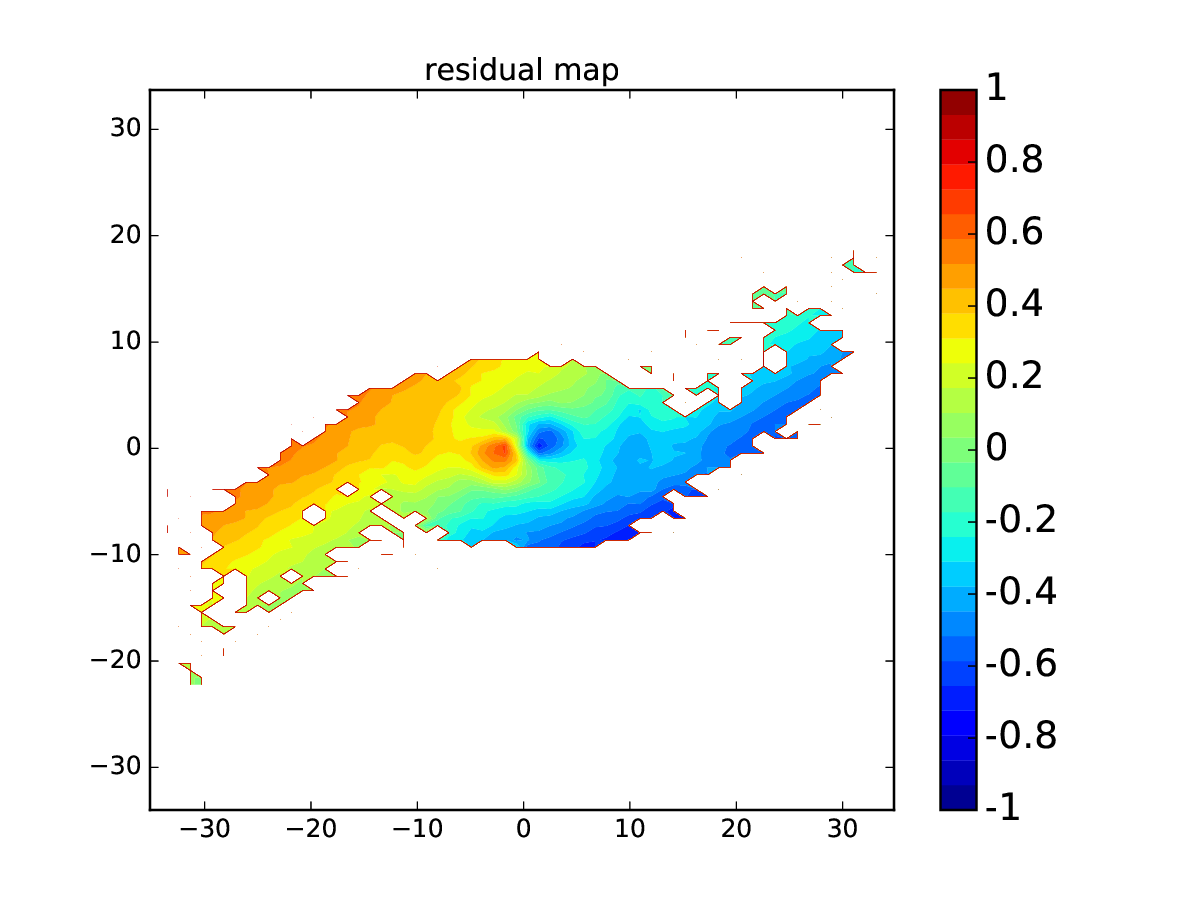}} 
\subfloat{\includegraphics[width = 2.in]{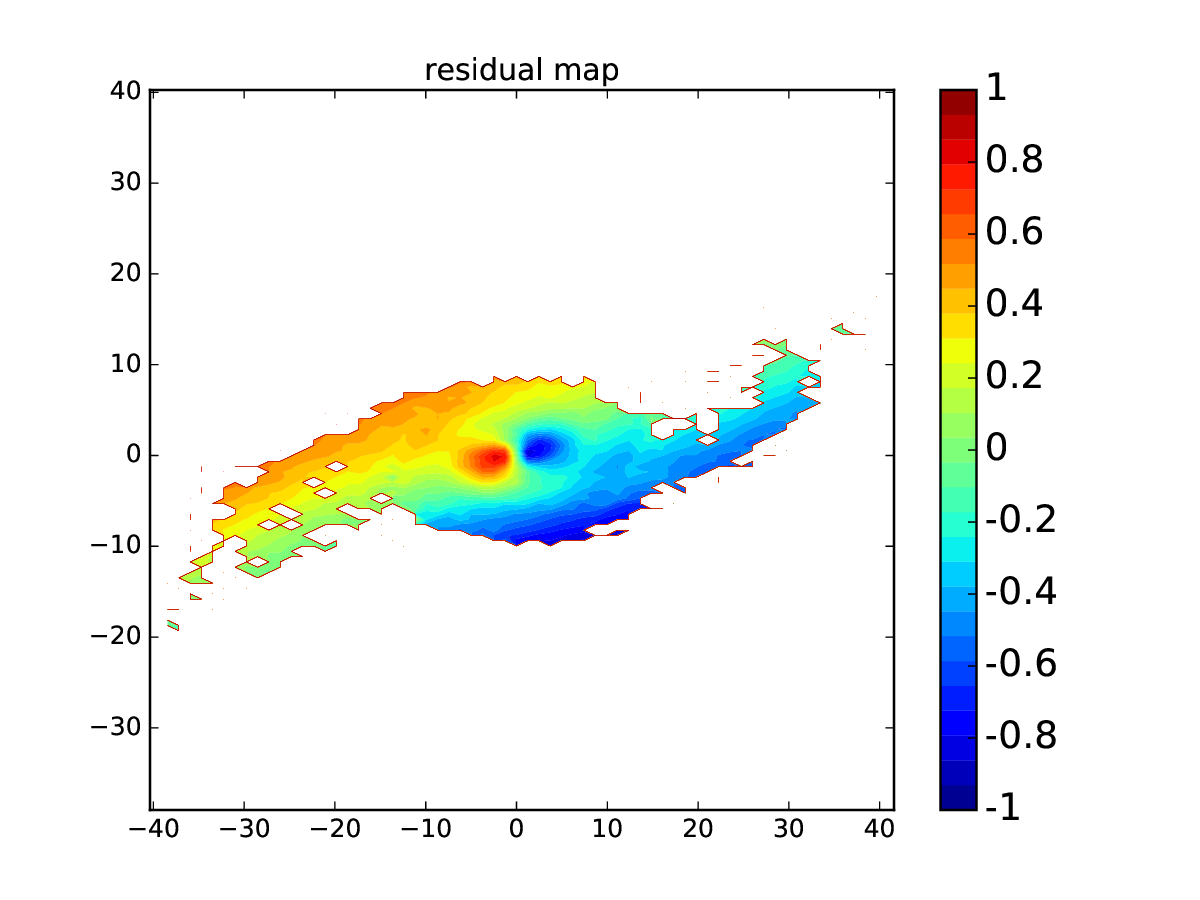}}
  \caption{Residuals for $i=60^\circ$ and different values of $j$ from $0^\circ$ to $80^\circ$ in steps of $10^\circ$.}
  \label{Fig-S9-i60res}
\end{figure*}

The reason for this surprising result is the strong correlation in
the alignment of the kinematic axis and the longest semi-principal 
axis of the
projected distribution, which {is a characteristic of our} 
out-of-equilibrium structures: as we have seen, the velocities in the outer
parts of the structure, which we are resolving in these mock
measurements, are radial and very preferentially oriented along the
axis, which is significantly elongated in the  structure. In
projection the major axis typically remains very close to this axis
--- other than for very specific observers, looking along the axis
with very small inclination angles --- and the large radial velocities
project out their components along this axis.

{A much more detailed and sophisticated analysis of
  observed projected velocity maps of spiral galaxies would evidently
  be required to establish or exclude their possible compatibility
  with velocity distributions qualitatively similar to those in our
  models, i.e. non-axisymmetric distributions with predominantly
  radial velocities very non-trivially correlated with the spatial
  distribution. As we have illustrated with our models, the motions in
  the outer parts of galaxies are in fact predominantly radial; }
there is no need to invoke a dark matter halo to explain them.
Indeed, as illustrated in the lower panels of Fig.~\ref{figure2} and
of Fig.~\ref{Fig-S9-i30j0a}, the mass estimate using the hypothesis of
rotational motions leads to an inferred mass that grows strongly with
radii, while the actual enclosed mass does not grow at all.

\subsection{{Flat Rotation Curves and Correlation between the Centripetal and Gravitational Acceleration}} 

{We conclude by speculating on two further important observational results about
velocity fields, and their possible explanation within the framework suggested by our models.}

{One of the noted properties of rotation curves of spiral galaxies
is that they are typically flat as a function of scale at the largest scales probed by observation,
although a great variety of behaviors are in fact observed in individual galaxies (see, e.g., \cite{Sofue_2017}).
Our models are not predictive in this respect:  we can obtain very different behaviors depending 
on the range of scale considered,  and notably whether we assume the region observed corresponds 
to  R2 or R3. Furthermore,
 the precise functional dependence on scale may be very different if
we modify, for example, the radial dependence of the initial angular velocity.
We note, however, that, if we consider the region R3, in which radial motions dominate, the
rotation curve (inferred by supposing the projected motions to arise from a rotating disk) 
will progressively flatten in time:  as the velocities are essentially ballistic the same
velocity range extends over a range of scale, which grows monotonically with time}.

{In this hypothesis of purely radial velocities with an approximately constant
 (i.e. very slowly increasing) amplitude, we note finally that one may also obtain very
 trivially in models like ours, also the observed phenomenological relation,  
 $a_c \propto \sqrt{g(r)}$, where $a_c$  is the centripetal acceleration 
inferred from the apparent motions, and $g(r)$ the gravitational acceleration of the 
visible baryonic mass (see, e.g., \cite{McGaugh_etal2016}), and 
which also underlies the so-called Modified Newtonian 
Dynamics \citep{Milgrom_1983,Milgrom_2016}.  Indeed, scale-independent
radial motions would give an inferred scale-independent rotation curve, 
and thus  $a_c \approx \frac{v_{max}^2}{r}$ where $v_{max}$ is the inferred
constant velocity of rotation, while $g(r) \approx GM_c/r^2$ where
$M_c$ is the mass in the virialized core.   
Thus, 
\[
a_c \approx \frac{v_{max}^2}{r} \approx \sqrt{a_0 \, g(r)}
\]
where 
\[
a_0= \frac{v_{max}^4}{GM_c} \;.
\]
The observed approximate constancy of $a_0$ for different systems then
corresponds to $v_{max}^4 \propto M_c$, i.e. the Tully-Fisher relation. }

\section{Discussion}
\label{discussion}

In summary, we have {shown, using simulations of 
evolution from very simple  toy initial conditions, 
that transient spiral-like structure may be generated in the far 
out-of-equilibrium evolution of a relaxing self-gravitating system. 
As will be detailed further in a forthcoming work 
\citep{benhaiem+joyce+syloslabini_2017}, the spatial 
organization in a spiral-like structure arises dynamically 
as particles that gain significant energy during an
initial collective contraction and expansion of the
system move outward, with the more energetic 
particles losing their transverse motion faster. 
The mechanism is completely different in nature from the 
usual perturbative mechanisms widely studied to explain 
such structure. Despite the unrealistically simplified
nature of the models, we have argued that a qualitative
comparison with observational data is possible:
our models show that the mechanism will 
generate structures with velocity fields
which have a very characteristic behavior.
This is a transition to predominantly radial 
motion with very small dispersion in the
outermost parts.  Surprisingly, we have found that
the projected motions of these regions can typically
be quite compatible with a rotating disk model,
up to residuals attributed to radial motion which
are very significant but of the order typically 
found in fitting rotating disk models to observations.}
This suggests
the possibility that these motions could be 
explained without invoking either dark matter or  a
modification of Newtonian gravity, which are
unavoidable if these
galaxies are modeled as stationary and rotating.  
Rather, these
motions might be consistent with the purely Newtonian gravitational
dynamics of the visible mass if the outer parts of the galaxy are far
from stationary and the motions are predominantly radial {and spatially
correlated in a non-axisymmetric distribution}, rather than
rotational. Instead of providing a single predictive model, we have
opened a Pandora's box of models, a different framework --- of
completely non-stationary mass distributions---  that must be compared 
 in much greater details with observations. Any such model
is obviously also very simplistic, {not just because of the
idealization of the initial conditions but also} in that it neglects everything 
but gravitational dynamics. Any detailed quantitative model will
of course necessarily need to consider more complex initial
conditions and also incorporate non-gravitational physics.  
{There are other obvious apparent shortcomings of the toy model.
For example, (i) spiral arms correspond to modest variations in mass 
density, and (ii) the time scale for collapse, as we have discussed, 
must be assumed short compared to the ages of old stars. The
former may plausibly be related to the low mass resolution
we have used, while the latter constraint may change in more
complex initial conditions.
Nevertheless we believe it is remarkable and tantalizing that the 
simple framework we have discussed produces structures 
bearing so much qualitative resemblance to astrophysical 
objects, and suggesting the possibility of a different and 
simple explanation for their observed projected motions.}

\begin{acknowledgments}
The authors of this  work were granted access to the HPC resources of The Institute for
Scientific Computing and Simulation financed by Region Ile de France
and the project Equip@Meso (reference ANR-10-EQPX- 29-01) overseen by
the French National Research Agency (ANR) as part of the
Investissements d'Avenir program.
\end{acknowledgments}

\newpage
\clearpage 

 %%%%%%%%%%%%%%%%%%%%%%%%%%%%%
 
\appendix %{Determination of apparent velocities} % (fold)
\label{projection}

{We detail here how  we construct the projected velocity maps
reported in in Sect.\ref{results} }from our simulated mass distributions. 
This projection is defined for a
random observer at infinity. It is convenient, in order to understand
the dependence on the orientation of the observer's LOS, to define
this orientation with respect to the principal axes of the mass
distribution.  Having done so, it then straightforward to determine
the projected velocities as a function of this orientation and the
components of the position and velocity in the principal axes.

%%%%%%%%%%%%%%%%%%%%%%%%%%%%%%%%%
\subsection*{A.1 Principal Axes} % (fold)
\label{sec:intertiamatrix}

We compute the inertia matrix of the mass distribution relative to an
origin taken at the minimum of the gravitational potential. We then
determine its eigenvalues ${\lambda}_i$, where ${\lambda}_1 \le
{\lambda}_2 \le {\lambda}_3$, and corresponding eigenvectors
$\vec{\lambda}_i$.  (The longest semi-principal axis } is then 
designated by a unit
vector $\vec{u}_1$ parallel to $\vec{\lambda}_1$, the intermediate
 semi-principal   axis
by a unit vector $\vec{u}_2$ parallel to $\vec{\lambda}_2$, and the
shortest  semi-principal axis by $\vec{\lambda}_3$. } The plane of 
the galaxy is then orthogonal to $\vec{\lambda}_3$.  We then rotate from our
original Cartesian axes (of the simulation) to determine the
components of the particle positions, $x_i$, and their velocities,
$v_i$, in the new basis $\{\vec{u}_i\}$.

\subsection*{A.2 Orientation of the Observer}

Following standard conventions (see, e.g., \cite{Backman_2004}) we
define the inclination angle $i$ of the observer as the angle between
his LOS and a vector orthogonal to the plane of the galaxy, {which
we take to be} $\vec{u}_3$.  Furthermore, as the galaxy is non-axisymmetric about this
axis, we define an azimuthal angle $j$ as the angle between the
projection into the galaxy plane of the LOS and the major axis.  Thus,
we write the unit vector parallel to the LOS as
\begin{equation}
\vec{u}_o= \sin(i) \cos(j) \,\vec{u}_1 + \sin(i) \sin(j) \,\vec{u}_2 +
\cos(i) \,\vec{u}_3\;,
\end{equation}

\subsection*{A.3 Determination of Projected Velocities}

To define the axes giving the observer's plane of projection it is
convenient first to define the set of axes
\bea && \vec{u}_x= \sin(j)
\,\vec{u}_1 - \cos(j) \,\vec{u}_2 , \\ \nonumber && \vec{u}_y= \cos(j)
\,\vec{u}_1 + \sin(j) \,\vec{u}_2 \;
\eea
in the plane of the galaxy. The vector $\vec{u}_y$ is thus parallel to
the axis of the projection in the plane of the galaxy of the observer
LOS, while $\vec{u}_x$ is the axis in the plane of the galaxy
orthogonal to the observer LOS.

The projected plane, orthogonal to the LOS, is then spanned by the
unit vectors
\begin{equation}
\vec{u}_{x^\prime}'= \vec{u}_x, \,\, \vec{u}_{y^\prime} = \vec{u}_o
\times \vec{u}_{x} \;.
\end{equation}
Using the expressions above, a little algebra gives
\begin{eqnarray}
&& \vec{u}_{1}= \sin(j) \vec{u}_{x^\prime} + \cos(j) \cos(i)
  \vec{u}_{y^\prime} + \cos(j) \sin(i) \vec{u}_o \\ &&\nonumber  \vec{u}_{2}=
  -\cos(j) \vec{u}_{x^\prime} + \sin(j) \cos(i) \vec{u}_{y^\prime} + \sin
  j \sin(i) \vec{u}_o \\ && \nonumber \vec{u}_{3}= -\sin(i) \vec{u}_{y^\prime} +
  \cos(i) \vec{u}_0 \;.
\end{eqnarray}
The position coordinates $(x^\prime, y^\prime)$ of the particles in
the plane of projection, and projected velocity $v_{los}=\vec{v}\cdot
\vec{u}_0$, can then be calculated, for any given observer $(i,j)$, as
\begin{eqnarray}
  \label{eq:proj}
  &&
  x^\prime =  x_1 \sin(j) - x_2 \cos(j)  \\
  && \nonumber 
y^\prime =  x_1 \cos(i) \cos(j) + x_2 \cos(i) \sin(j) - x_3 \sin(i)  \\
  && \nonumber 
v_{los} =   v_1 \sin(i) \cos(j) + v_2 \sin(i) \sin(j) + v_3 \cos(i)  \;. 
\end{eqnarray}

%%%%%%%%%%%%%%%%%%%%%%%%%%%%%%%

\subsection*{A.4 One-dimensional Apparent Velocity Profiles}  
\label{sec:section_1dLos}

{Most observations of apparent velocities are not fully
  two-dimensional, but given along a specific axis (corresponding to
  the orientation of the slit used for the spectographic
  measurements). In order to obtain such one-dimensional velocity
  profiles we define two such slits: one aligned parallel to the kinematic
  axis, i.e. the axis along which there is the maximum gradient of the
  LOS velocity (details below), and one orthogonal to this
  direction. We have also considered projections along the 
 major axis and minor axis of the projected distribution (defined following a
  procedure analogous to that described above for the three-dimensional case).  
  The slit is {assumed} to be rectangular, of a width $\Delta$
  which is a small fraction of the minor axis.}  From these LOS
velocity profiles along the kinematic axis $v_{los}(R)$, we estimate
the mass $M_c(R)$ enclosed in the radius $R$ assuming that particles
are in stationary circular orbits as
\be 
\label{mcr}
M_c(R) = \frac{v_{los}^2(R) R}{\sin(i) G} \;,
\ee
where the inclination angle is estimated from the projection as
described below.

\subsection*{A.5 Velocities for Rotating Disk Model} 
\label{sec:rotationnal_model}

If one models a galaxy as a disk, the projected LOS velocities can be
written (see, e.g., \cite{Backman_2004}) as \be
\label{eq:beckman} 
v_{los} (r, \phi) = v_\theta \sin(i) \cos (\theta) + v_R \sin (i) \sin
(\theta) \;, \ee where $(r, \phi)$ are polar coordinates in the plane
of the projection, with $\phi$ defined relative to the axis orthogonal
to the observer LOS (i.e. parallel to the axis $\vec{u}_x$ defined
above), and $v_\theta$ and $v_R$ are the components of the velocity
field given in polar coordinates $(R, \theta)$ in the plane of the
galaxy (with $\theta$ defined relative to the same axis $\vec{u}_x$,
which is also in the plane of the galaxy). The polar coordinates are
related by the transformation
\bea
\label{eq:beckman-tran} 
&&
\tan (\theta) = \tan (\phi) / \cos(i)
\\ \nonumber &&
R=r \cos (\phi) / \cos (\theta) \;. 
\eea
For a purely rotating axisymmetric model, $v_R=0$
and $v_\theta= v_\theta (R)$. The kinematic axis
is that along which there is maximal variation
of the projected velocity, i.e. $\theta=\phi=0$.

\subsection*{A.6 Fitting to a Rotating Disk Model} 

To fit our projected velocity data to a rotating axisymmetric disk we first
estimate  from our data the orientation of the  kinematic axis. {We determine the kinematic
  axis as the axis passing through the center of mass of the
  distribution and along which the difference of the velocities at the
  two extreme points is maximal.}

While this axis must strictly be the major axis of the projection if
the underlying distribution is really a disk, this is generally not
the case for our distributions that are not axisymmetric. However,
because in our models the directions of the radial velocities are
strongly correlated with the real three-dimensional major axis of the
non-axisymmetric distribution (see below), the offset between the
kinematic axis and the projected major axis is, in fact, typically
(i.e. for a large fraction of random observers) quite small.  Such
offsets are, indeed, typically seen in observations (see,
e.g., \cite{Erroz-Ferrer_etal_2015}).

To find the best-fitting rotating disk model, we need to determine the
inclination angle $i$: we do this by minimizing the residuals between
the rotational model, computed for a generic $i$, and the actual data
on each grid cell. To do so we compute first, for each grid cell,
{labeled by $\alpha$} and centered on projected coordinates
$x_\alpha^\prime,y_\alpha^\prime$, the polar coordinates as defined
above:
\begin{eqnarray}
  r_\alpha &=& \sqrt{(x_\alpha^\prime)^2+(y_\alpha^\prime)^2} \\ \nonumber 
  \phi_\alpha &=& \arccos(x_\alpha^\prime/r_\alpha) \\ \nonumber 
  R_\alpha &=& r_\alpha \sqrt{\cos(\phi_\alpha)^2+
    \sin(\phi_\alpha)^2/\cos(i)^2} \\ \nonumber  
  \theta_\alpha &=& \arctan(\tan(\phi_\alpha)/\cos(i)) \;. 
\label{unproj-proj} 
\end{eqnarray}
Then, for the given value of the inclination angle $i$, we use
Eq.\ref{eq:beckman} (with $v_R=0$) to compute the LOS velocity of the
rotational model, denoted $v_{los,model}^\alpha$.
Note that in the case where the unprojected size of the galaxy is
larger than the maximum distance at which the LOS velocity profile
extends, we perform a linear fit over the last five points of
$v_{los}(R)$ and then extrapolate using this fit to a higher radius.
Finally, in order to get the best-fitting inclination angle, we
minimize the sum of the residuals in all the cells with respect to
$i$, i.e.
\be
\mbox{Residuals} = \sum_{\alpha} |v_{los}^{\alpha} - v_{los,model}^{\alpha}| \;. 
\ee  
%
%%%%%%%%%%%%%%%%%%%%%%%%%%%%%%%%%%%%%%5
%%%%%%%%%%%%%%%%%%%%%%%%%%%%%%%%%%%%%%%
%%%%%%%%%%%%%%%%%%%%%%%%%%%%%%%%%%%%%%%5
 
 %%%%%%%%%%%%%%%%%%%%%%%%%%%%%

%\bibliography{bibliography}{bibliography.bib}

\end{document}